\documentclass[11pt,fleqn]{article}
\usepackage{amsmath,amssymb,graphicx,epsfig}

\def \lsim{\mathrel{\vcenter
{\hbox{$<$}\nointerlineskip\hbox{$\sim$}}}}

\newcommand{\Tr}{\text{Tr}}

\def\bea{\begin{eqnarray}}
\def\eea{\end{eqnarray}}
\def\be{\begin{equation}}
\def\ee{\end{equation}}
\def\ba{\begin{array}}
\def\ea{\end{array}}
\def\nn{\nonumber}
\def\a{& \hspace{-11pt}}
\def\b{& \hspace{-7pt}}

\font\tenrsfs=rsfs10
\font\sevenrsfs=rsfs7
\font\fiversfs=rsfs5
\newfam\rsfsfam
\textfont\rsfsfam=\tenrsfs
\scriptfont\rsfsfam=\sevenrsfs
\scriptscriptfont\rsfsfam=\fiversfs
\def\mathscr#1{{\fam\rsfsfam\relax#1}}

\begin{document}

\thispagestyle{empty}

\begin{center}

$\;$

\vspace{1cm}

{\huge \bf Mildly sequestered supergravity models \\[2mm]
and their realization in string theory}

\vspace{1.5cm}

{\Large {\bf Christopher Andrey} and {\bf Claudio~A.~Scrucca}}\\[2mm] 

\vspace{0.8cm}

{\large \em Institut de Th\'eorie des Ph\'enom\`enes Physiques\\ 
Ecole Polytechnique F\'ed\'erale de Lausanne\\ 
CH-1015 Lausanne, Switzerland\\}

\vspace{0.2cm}

\end{center}

\vspace{1cm}

\centerline{\bf \large Abstract}
\begin{quote}

We elaborate on the idea that five-dimensional models where sequestering 
is spoiled due to contact interactions induced by vector multiplets may still 
be mildly sequestered if a global version of the gauge symmetry associated 
to the latter survives in the hidden sector. Interestingly, it has been argued that 
although in such a situation non-trivial current-current contact interactions are 
induced by the heavy vector modes, these do not induce soft scalar masses, as 
a consequence of the global symmetry. We perform a detailed study of how this 
hybrid mechanism can be implemented in supergravity and string models, focusing 
on the prototypical case of heterotic M-theory orbifolds. We emphasize that 
in general the mechanism works only up to subleading effects suppressed 
by the ratio between the global symmetry breaking scale in the hidden sector 
and the vector mass scale or the Planck scale. We also argue that this mild 
sequestering mechanism allows to rehabilitate the scenario of dilaton domination 
of supersymmetry breaking, which is incompatible with dilaton stabilization 
in its original version, by exploiting the fact that hidden brane fields do 
contribute to the cosmological constant but not to soft terms, thanks to the 
global symmetry.

\vspace{5pt}
\end{quote}

\renewcommand{\theequation}{\thesection.\arabic{equation}}

\newpage

\section{Introduction}
\setcounter{equation}{0}

One of the most important issues in models with high-scale supersymmetry 
breaking, like most notably string models in their traditional conception, is 
the supersymmetric flavor problem. Scalar soft masses, in particular, should 
be approximately universal or suitably aligned to avoid excessively large 
contributions to certain flavor changing processes. 
Several solutions to this problem have been proposed and concern 
specific features of the higher-dimensional contact operators mixing visible 
and hidden sector fields that induce sfermion masses after spontaneous 
supersymmetry breaking. 

One natural possibility is provided by flavor symmetries, of the same type 
as those designed to explain the structure of Yukawa couplings. Such 
symmetries may also be used to suitably constrain the contact terms 
giving origin to scalar masses, in such a way to suppress the off-diagonal 
entries of the soft scalar mass matrices \cite{LNSflavor,NSflavor}.
A different possibility is sequestering along an extra dimension \cite{seq}, 
where the local contact terms are absent due to the geometric separation 
between the visible and the hidden sectors. In such a situation, which 
corresponds to the so-called no-scale models \cite{noscale}, the 
scalar masses vanish at the classical level and are induced only by 
approximately flavor-universal loop effects of various types, like 
for instance anomaly mediation \cite{seq,anomed}, radion-mediation 
\cite{GR} or brane-to-brane mediation \cite{RSS,BGGLLNP}.
A third possibility is that of conformal sequestering by large anomalous 
dimensions, where the contact terms are suppressed by some strong 
renormalization group effects from the hidden sector \cite{confseq} or 
the visible sector \cite{NS}.

In the particular context of string models, one may also rely on 
the specific structure taken by the contact terms and imagine situations
where the soft terms turn out to be approximately universal. For instance,
the soft scalar masses originating from supersymmetry breaking in the 
neutral moduli sector can be characterized in a rather model-independent
way \cite{KL,BIM}. One can then argue that if the dilaton were dominating 
supersymmetry breaking, the soft terms would be approximately universal, 
as a consequence of the universality of the direct couplings of the dilaton,
with flavor-violating effects emerging only at the loop level and being thus 
barely sufficiently suppressed \cite{LN}. Unfortunately, it turns out that 
the dilaton dominated scenario is incompatible with dilaton stabilization, 
under the assumption of weak string coupling \cite{C,BdA,GRS}. On the other 
hand, a non-trivial contribution from the geometric moduli would allow for 
stabilization of all the fields, but would generically spoil universality, unless 
the modular weights parametrizing their couplings to visible fields enjoy some
special properties. In addition, whenever there occurs a spontaneously broken 
extra gauge symmetry, additional contributions to soft terms are induced when 
the corresponding heavy vector multiplets are integrated out, and these are 
also not universal, unless the charges of the matter fields are universal 
\cite{KK,ADM, ChoiDFF, Scrucca}.

The idea of sequestering naturally fits into the framework of string/M-theory,
since this automatically provides extra dimensions and localized sectors.
However, it has been argued in \cite{noseq} that there is an endemic difficulty 
against realizing even effectively the minimal 5D setup 
proposed in \cite{seq} in concrete string models, due to the generic 
appearance of vector multiplets in the 5D bulk. From the 
4D point of view, these lead to a chiral multiplet zero mode 
associated to a new modulus, plus heavy Kaluza--Klein vector multiplet 
modes which, when integrated out, produce  non-trivial contact terms between
the visible and hidden sectors already at the classical level. This phenomenon
occurs in a rather clear way in the case of heterotic M-theory compactified on 
a Calabi-Yau or an orbifold. In that situation, any non-minimal K\"ahler modulus 
arising in addition to the overall volume modulus is associated to a 5D
vector multiplet, whose heavy KK modes induce non-trivial corrections to the 
effective K\"ahler potential. As a result, the simple maximally symmetric scalar 
manifold arising in the minimal case with one K\"ahler modulus and displaying 
sequestering is changed to a less symmetric scalar manifold where sequestering 
is spoiled as soon as additional K\"ahler moduli occur. In such a situation, one 
then generically finds that non-vanishing and non-universal soft scalar masses 
are generated out of the contact terms induced by the vector multiplets.

Interestingly, it has been proposed that the spoiled sequestering of generic string 
models can be rehabilitated in a milder form by making additional assumptions 
concerning the symmetries of the hidden sector \cite{KMS}. 
The basic point behind this idea was explained in \cite{SS}, and relies on the fact 
that although the contact terms that occur do not vanish, they have a very particular 
form, due to the fact that they are induced by integrating out a heavy vector superfield. 
More precisely, they essentially consist of the product of two current superfields $J$, 
associated to the symmetry that was gauged by the heavy vector superfields, 
divided by the square of the mass scale $M$ of these modes. In the 
low-energy effective theory, there may then remain a global version of the 
original symmetry, implying the conservation of these current superfields: $D^2 J = 0$. 
This Ward identity implies not only that the $J^\mu$ component of $J$ is conserved, 
but also that its $F$ and $D$ components vanish. It then follows that the superfield 
contact operator $J M^{-2} J$ gives vanishing soft scalar masses.
This mildly sequestered situation, where non-trivial contact terms arise 
but do not give any contribution to soft scalar masses due to some global 
symmetry of the hidden sector, can be implemented more generically 
in string models \cite{KMS}. 

The aim of this paper is to make a more detailed investigation of how 
this mechanism of mild sequestering can be implement in supergravity 
and string models, focusing on the prototypical case of heterotic M-theory
orbifolds. We will display very explicitly how the contact terms spoiling 
sequestering can be recast into the form of current-current interactions in 
these models, by rederiving the 4D effective K\"ahler potential
from a 5D intermediate starting point and explicitly integrating 
out the heavy vector multiplets at the superfield level. We will also examine 
more closely the effects of the spontaneous breaking of the assumed global 
symmetry, which is necessary to have non-vanishing VEVs for the hidden
sector auxiliary fields, taking into account the presence of gravity.
We will argue that the mechanism of mild sequestering actually works only 
up to subleading effects suppressed by the ratio of the scale of global 
symmetry breaking in the hidden sector and the vector mass scale or the 
Planck scale. We will finally point out that this mild sequestering mechanism 
allows to consistently realize a generalized version of the dilaton domination 
scenario, by exploiting the fact that hidden brane fields do contribute to the 
cosmological constant but not to soft masses, thanks to the global symmetry.

\section{Mild sequestering from global symmetries}
\setcounter{equation}{0}

Let us begin by illustrating in more detail the general ideas of \cite{SS,KMS}, 
by working first in rigid supersymmetry. The starting point is to consider a situation 
where the effective K\"ahler potential contains only a very special kind of contact 
terms mixing visible and hidden sector fields, which are induced at the classical 
level by the exchange of heavy vector multiplets. To evaluate the structure of 
such an effect, let us then denote by $J_a$ the current superfield acting as 
linear source for the heavy vector superfield $V^a$, and by $M^{ab}$ the 
mass controlling the quadratic potential for $V^a$. The K\"ahler potential can 
then be expanded in powers of $V$ as 
$K^{\rm mic} \simeq K + J_a V^a + \frac 12 M^2_{ab} V^a V^b$, whereas 
$W^{\rm mic} = W$. At this point, the field $V^a$ can be integrated out by neglecting space-time 
derivatives in its equation of motion, which becomes $K^{\rm mic}_a = 0$ and implies
$V^a \simeq - M^{\mbox{-}2\,ab} J_b$. Plugging back this solution, one finds that the 
effective K\"ahler potential is given by $K^{\rm eff} \simeq K - \frac 12 J_a \,M^{\mbox{-}2\,ab} J_b$,
whereas the superpotential is not affected and $W^{\rm eff} = W$. 
We see then that even if the visible and the hidden sectors did not mix 
in $K=K^{\rm v} + K^{\rm h}$, such a mixing is induced by the exchange of $V^a$ 
through the contact term involving $J_a = J_a^{\rm v} + J_a^{\rm h}$, which contains
\be
K^{\rm eff}_{\rm cont} \simeq - J^{\rm v}_a \,M^{\mbox{-}2\,ab} J^{\rm h}_b \,.
\label{contact}
\ee
When the hidden sector superfields get a non-vanishing VEV for their
auxiliary fields, this generically induces soft scalar masses for the visible sector 
superfields. There is a direct effect coming from the $D$ 
component of $J_a^{\rm h}$, and an indirect effect coming from its $F$ component. 
In principle, there are also similar effects coming from the $D$ and $F$ components 
of $M^{\mbox{-}2ab}$, whenever this depends on the visible and hidden sector 
fields. However, these effects are on the same footing as those that we already 
discarded by expanding $K$ in powers of $V^a$. They involve additional powers 
of the ratio between the VEVs of the hidden sector scalars and the heavy mass scale, 
which we may assume to be small. They are thus less important, and we shall neglect 
them for the time being.

The main observation of \cite{SS,KMS} is that there is one particularly simple
situation where a contact term of the form (\ref{contact}) does in fact not induce 
any soft scalar mass. This is when the hidden sector possesses a global 
symmetry implying the conservation of the current $J_a^{\rm h}$, which at 
the level of superfields means:
\be
D^2 J_a^{\rm h} = 0 \,.
\ee
Indeed, at the component level this implies not only the conservation of the 
$\theta \sigma^\mu \bar \theta$ component of $J_a^{\rm h}$, that is
$\partial_\mu J^{{\rm h}\mu}_a = 0$ as dictated by N\"other's theorem, 
but also the vanishing of its $\theta^2$ and $\theta^2 \bar \theta^2$ components:
\be
J_a^{\rm h}|_F = 0 \,,\;\; J_a^{\rm h}|_D = 0\,.
\ee
As a result of these Ward identities, the most important contributions to soft scalar masses 
from (\ref{contact}) disappear. One can then conclude that the global symmetry forces 
the soft scalar masses to vanish, in first approximation:
\be
m^2_{\alpha \bar \beta} \simeq 0 \,.
\ee

Notice that the subleading effects that we have mentioned above and discarded would 
in general give a non-trivial contribution to these scalar masses. However, this is suppressed by 
some power of the ratio $v^2/M^2$, where $v$ is the scale of spontaneous breaking 
of the global symmetry defined by the VEVs of the hidden sector scalar fields, more 
properly defined as:
\be
\epsilon^{{\rm gau}\,a} \sim M^{\mbox{-}2\,ab} v^2_b \,.
\label{egau}
\ee
One can then assume that this parameter is small and neglect this effect, as already said. 
This is reasonable, since the breaking scale $v$ is a priori arbitrary and unrelated to $M$, 
although one should keep in mind that the emergence of non-trivial VEVs for the hidden sector 
auxiliary fields implies that it is non-vanishing. 

The above mechanism can be rephrased more intuitively as follows. In the microscopic 
theory, possible soft scalar masses can come only from the coupling between the visible sector 
current $J_a^{\rm v}$ and the vector superfield $V^a$, after the latter gets a 
non-trivial VEV for its auxiliary field from the interaction with the hidden sector current 
$J_a^{\rm h}$. This $D$-term breaking interpretation in the microscopic theory is perfectly
equivalent to the $F$-term breaking picture obtained in the effective theory, as a consequence
of the fact that on-shell the $D^a$ auxiliary field of the vector multiplet is determined in 
terms of the $F^i$ auxiliary fields of the hidden sector (see for example \cite{ADM}).
The effect of the global symmetry is then to force $D^a$ to vanish, as a consequence
of the relations that it implies among the various $F^i$ in the hidden sector.

The mechanism by which the global symmetry constrains the values of the auxiliary 
fields can be made more transparent by considering more explicitly the general case 
of a theory that is invariant under some global symmetry acting as $\delta_a \Phi^i = k_a^i(\Phi)$
on the superfields, in terms of some holomorphic Killing vectors $k_a^i$.
The Lagrangian is then invariant if $\delta_a K = f_a + \bar f_a$ and $\delta_a W = 0$,
where $f_a$ is a holomorphic function parametrizing a K\"ahler transformation.
In such a situation, the N\"other current takes the following form:
\be
J_a = {\rm Im} \big(k_a^{i} K_i - f_a\big) \,. 
\label{currentsup}
\ee 
Using the equations of motion, which read $- \frac 14 \bar D^2 K_i + W_i = 0$, and the 
almost invariance of $K$ and $W$, which imply respectively that ${\rm Re}(k_a^{i} K_i - f_a) = 0$
and $k_a^i W_i = 0$, it is straightforward to verify that this current indeed satisfies the 
conservation law
\be
D^2 J_a = 0 \,.
\ee
As already said, this implies in particular that the $F$ and $D$ components of $J_a$ vanish.
More explicitly, after using the invariance of $K$, these informations become:
\be
\bar k_{ai} F^i = 0 \,,\;\; \nabla_{i} k_{a \bar \jmath} \,F^i F^{\bar \jmath} = 0 \,.
\label{wardrig}
\ee
These two relations, where $\nabla_i$ denotes the covariant derivative on the K\"ahler 
manifold spanned by the scalar fields, can be easily verified also in a more direct way
using component fields. To do so, recall that the stationarity condition reads $\bar W^j \nabla_i W_j = 0$. 
One also has $F^i = - \bar W^i$. Then, the condition of invariance of $W$ leads directly to 
$\bar k_{a i} F^i = 0$, whereas acting on this invariance condition with $\bar W^i \nabla_i$ 
and using the stationarity condition, one deduces that 
$\nabla_{i} k_{a \bar \jmath} \, F^i F^{\bar \jmath} = 0$. 

In local supersymmetry, on the other hand, the situation is slightly more subtle. One can presumably 
define a superfield current in the superconformal formalism, but we will not attempt to do so. 
Rather, we shall derive the generalization of the two linear and quadratic relations (\ref{wardrig})
on the auxiliary fields by proceeding in a direct way, in components. In order for the theory 
to be invariant under global transformations of the form $\delta_a \Phi^i = k_a^i (\Phi)$, we must 
now require that  $\delta_a K = f_a + \bar f_a$ and $\delta_a W = e^{-f_a} W$, where $f_a$ is a 
holomorphic function parametrizing a K\"ahler transformation. This means that the function 
$G = K + {\rm log}|W|^2$ must be invariant, $\delta_a G = 0$.
The computation yields a result that shows that the conservation laws are altered by supergravity 
effects proportional to the gravitino mass. More precisely, assuming vanishing cosmological constant
one finds
\be
\bar k_{a i} F^i = - i D_a m_{3/2} \,,\;\; 
\nabla_{i} k_{a \bar \jmath} \, F^i F^{\bar \jmath} = - 2i D_a m_{3/2}^2 \,,
\label{wardsugr}
\ee
where:
\be
D_a = \frac {{\rm Im} (k_a^i F_i)}{m_{3/2}} \,.
\label{Da}
\ee
The notation for this last quantity is reminiscent from the fact that if the global symmetries 
generated by $k_a^{i}$ were gauged by light vector multiplets $V^a$ in the hidden sector, 
$D_a$ would determine the value of the auxiliary fields of these extra vector multiplets, 
after multiplication by the gauge coupling matrix $h^{ab}$: $D^a = h^{ab}D_b$. 
To check the above relations, one can proceed along the same lines as in the rigid case. 
Recall first that the invariance of $G$ implies ${\rm Re}(k_a^{i} G_i) = 0$, the vanishing 
of the cosmological constant implies that $G_i G^i = 3$ and the stationarity condition 
reads $G^j \nabla_i G_j + G_i = 0$. Recall also that  $F^i = - e^{G/2} G^i$ 
and $m_{3/2} = e^{G/2}$. Then, the condition of invariance of $G$ directly leads to 
$\bar k_{ai} F^i = i\,{\rm Im} (k_a^{i} G_i) \, m_{3/2} $, whereas contracting the stationarity 
condition with $k_a^{i}$ and using the invariance condition of $G$ as well as its derivatives, 
plus the vanishing cosmological constant condition, one deduces after a straightforward 
computation that $\nabla_{i} k_{a \bar \jmath} \, F^i F^{\bar \jmath} = 2 i \, {\rm Im} (k_a^{i} G_i) \,m_{3/2}^2$. 

We see that in the presence of gravity, the Ward identities that are relevant for the mechanism 
of mild sequestering get modified. More precisely, restoring explicitly the dependence 
on the Planck scale $M_{\rm Pl}$, and recalling that the condition of vanishing cosmological 
constant implies that $|F^i| \lsim m_{3/2} M_{\rm Pl}$, we see that the new effects are 
suppressed by powers of the following dimensionless parameter:
\be
\epsilon_a^{\rm gra} = \frac {D_a}{M_{\rm Pl}^2} \,.
\label{egra}
\ee
As a result of these effects, the global symmetry in the hidden sector does not imply
any longer that the scalar masses vanish, but rather that they are suppressed by 
some power of the above parameter. But again, this can be reasonably assumed 
to be small, and these effects can then be neglected. Actually, it is not totally clear 
whether it would make sense at all to keep the above non-trivial gravitational effects. 
Indeed, if for some reason one were allowed to count the quantity $D_a$ as scaling 
proportionally to some power of $m_{3/2}$, the above effect would have to be neglected, 
since it would be of the same order as effects coming from higher-derivative 
terms in the gravity sector. This counting is to some extent suggested by the fact 
that the relations (\ref{wardsugr}) can be extended to theories where the global 
symmetry is gauged and $D^a = h^{ab} D_b$ corresponds to the auxiliary field of 
the vector field introduced for this gauging. More precisely, the first relation stays 
unchanged, and the second receives on the right hand side two extra contributions 
that are linear and quadratic in $D^a$ (see for example \cite{Scrucca}), which 
disappear when the gauge coupling is switched off.
In this situation, the vanishing of the cosmological constant implies that 
$|D_a| \lsim g_a^{-1} m_{3/2} M_{\rm Pl}$, where $g_a$ represents the diagonalized 
gauge couplings. For finite $g_a$, we see that the two corrections appearing in the right 
hand sides of (\ref{wardsugr}) would then be subleading in the number of auxiliary fields, 
since $m_{3/2}$ can be assimilated to the gravitational scalar auxiliary 
field.\footnote{See \cite{BGS} for a general discussion of this issue.}
For $g_a \to 0$, on the other hand, this argument does no longer hold true, 
and one must in principle keep these corrections.

In this paper, we shall be primarily concerned with situations where the global 
symmetry responsible for the mild sequestering mechanism is linearly realized. 
Let us then see more specifically how things work in that case. Omitting the 
indices, we may consider some group $G$ with generators $\lambda_a$, and 
take Killing vectors of the form $k_a = i \lambda_a \Phi$. On the other hand, 
$K$ and $W$ can be generic real and holomorphic non-linear functions that 
are invariant under the transformations $\delta_a \Phi = k_a$. The minimal 
possibility for the microscopic theory is then that $K = \bar \Phi \Phi$ and 
$J_a = \bar \Phi \lambda_a \Phi$. After integrating out the heavy vector fields,
one obtains a more complicated effective K\"ahler potential of the approximate
form $K^{\rm eff} \simeq K - \frac 12 J_a M^{\mbox{-}2\,ab} J_b$, displaying 
mild sequestering. One of our goals will be to study more precisely the structure 
of the additional subleading corrections to this result, which as discussed are 
both of gauge and gravitational nature and parametrized by (\ref{egau}) and 
(\ref{egra}). To do this, we shall rely only on the Ward identities (\ref{wardsugr}),
which are exact and have been derived using supergravity component fields.
We shall moreover focus on a particular class of models where the effective 
theory is simple enough to allow an exact study.

\section{Non-sequestered string models}
\setcounter{equation}{0}

The sequestered model considered in \cite{seq} arises from a very minimal
5D supergravity theory compactified on $S^1/Z_2$, in which only the gravitational 
multiplet propagates in the bulk and matter multiplets are confined to 4D 
hyper-planes corresponding to the $Z_2$ fixed-points. In trying to effectively 
implement this model within string theory, one realizes however that such a minimal 
situation is rather unnatural, and one typically finds additional hyper and vector 
multiplets in the bulk \cite{noseq}.\footnote{The same situation arises also as soon 
as one tries to start from a theory in dimension higher than five. See for instance \cite{FLL} for 
a 6D example.} A prototypical class of such models is obtained
by considering 11D M-theory compactified on an orbifold of the type 
$T^6/\Gamma \times S^1/Z_2$ \cite{HW}. Below the compactification scales, this 
yields a 4D effective theory. However, if the volume of $S^1/Z_2$ is much larger 
than the volume of $T^6/\Gamma$, one may also study the intermediate 5D 
effective theory obtained by compactifying on $T^6/\Gamma$, and then reobtain 
the 4D effective theory by further compactifying on $S^1/Z_2$. One then finds a 
natural generalization of the minimal 5D models displaying sequestering, with 
some additional dynamics in the bulk, whose details depend on $\Gamma$.

Let us review the main features of this kind of theories, focusing on those points 
that will be directly relevant for our purposes. The starting point is 11D supergravity, 
whose bosonic fields consist of the metric $g_{AB}$ and a three-index antisymmetric 
tensor $C_{ABC}$. Upon compactification on a 6D internal manifold, these fields split 
as follows:  $g_{AB} \rightarrow g_{MN}, g_{Mn}, g_{mn}$, 
$C_{ABC} \rightarrow C_{MNP}, C_{MNp}, C_{Mnp}, C_{mnp}$.
When the internal manifold has a non-trivial $SU(3)$ holonomy, the internal rotation
symmetry is broken at least as $SO(6) \rightarrow U(1) \times SU(3)$. One gets 
then a 5D theory with minimal supersymmetry, a $U(1)$ $R$-symmetry and 
an $SU(3)$ flavor symmetry. Finally, this $SU(3)$ flavor symmetry can be further 
reduced to a subgroup $G$, depending on the model, but for Abelian orbifolds 
$G$ has the same rank as $SU(3)$.

To discuss the field content of the 5D theory, it is convenient to describe the 
internal orbifold with three complex coordinates and distinguish the fields 
in terms of the internal symmetry $SO(6) \to U(1) \times SU(3)$. 
From $g_{AB}$ we get the following fields: $g_{MN}$ gives $1$ symmetric tensor
in the ${\bf 1}$, $g_{Mm} \Leftrightarrow g_{Mi}, g_{M\bar \imath}$ 
gives $6$ vectors in the ${\bf 6} \to {\bf 3} \oplus {\bf \bar 3}$, and 
$g_{mn} \Leftrightarrow g_{i \bar \jmath}, g_{ij}, g_{\bar \imath \bar \jmath}$ 
gives $21$ scalars in the ${\bf 21} \to {\bf 1} \oplus {\bf 8} \oplus {\bf 6} \oplus {\bf \bar 6}$.
From $C_{ABC}$ we get instead the following fields: $C_{MNP}$ gives 
by dualization $1$ scalar in the ${\bf 1}$, 
$C_{MNp} \Leftrightarrow C_{MNi}, C_{MN\bar \imath}$ gives by dualization 
$6$ vectors in the ${\bf 6} \to {\bf 3} \oplus {\bf \bar 3}$, 
$C_{Mnp} \Leftrightarrow C_{M i \bar \jmath}, C_{M ij}, C_{M\bar \imath \bar \jmath}$ 
gives $15$ vectors in the ${\bf 15} \to {\bf 1} \oplus {\bf 8} \oplus {\bf 3} \oplus {\bf \bar 3}$,
and finally $C_{mnp} \Leftrightarrow C_{i j k}, C_{\bar \imath \bar \jmath \bar k},C_{i j \bar k}, C_{i \bar \jmath \bar k}$ 
gives $20$ scalars in the ${\bf 20} \to {\bf 1} \oplus {\bf 1} \oplus {\bf 3} \oplus {\bf \bar 3} \oplus {\bf 6} \oplus {\bf \bar 6}$.
However, the orbifold projection kills a subset of these states, and also breaks 
$SU(3) \to G$. More precisely, it keeps the ${\bf 1}$, it kills all the states of the ${\bf 3}$, 
and keeps $h_{1,1}-1$ states of the ${\bf 8}$ and $h_{2,1}$ states 
of the ${\bf 6}$, with $h_{1,1}$ and $h_{2,1}$ depending on the orbifold action. In total,
we thus get $1$ symmetric tensor, $h_{1,1}$ vectors and $h_{1,1} + 4\,h_{2,1} + 3$ 
scalars. This is the bosonic field content of 5D supergravity with 
$h_{1,1} - 1$ vector multiplets and $h_{2,1} + 1$ hypermultiplets
\cite{CCDF,AFT,LOSW}. Moreover, we see that the $h_{1,1}-1$ vector multiplets 
transform in some representation ${\bf h_{1,1} \!-\!1}$ of $G$ arising from the 
projection of the ${\bf 8}$ of $SU(3)$, and similarly the $h_{2,1} + 1$ hypermultiplets 
transform in some (reducible) representation ${\bf 1} \oplus {\bf h_{2,1}}$ of $G$ arising 
from the projection of the $ {\bf 1} \oplus {\bf 6}$ of $SU(3)$. 

When further compactifying on $S^1/Z_2$, one obtains a 4D supergravity theory with minimal 
supersymmetry. The structure of multiplets can be understood by first recasting the multiplets of the 
5D theory as multiplets of $N=2$ supersymmetry in 4D, and then figuring out their 
content in terms of $N=1$ multiplets with definite $Z_2$ parities. 
The universal gravitational multiplet decomposes as ${\cal G} = (E,T^0;\Psi)$, 
where the even components are the gravitational multiplet $E$ plus a chiral multiplet $T^0$, 
and the odd component is some more complicated multiplet $\Psi$. The $h_{1,1}-1$
non-universal vector multiplets decompose as ${\cal V}^a = (T^a;V^a)$, where the 
even components are chiral multiplets $T^a$ and the odd components are vector multiplets 
$V^a$. Finally, the universal hypermultiplet is decomposed as ${\cal H} = (S;S^c)$, where 
the even component $S$ is a chiral multiplet and the odd component $S^c$ too, and 
similarly the $h_{2,1}$ non-universal hypermultiplets decompose as ${\cal H}^x = (Z^x;Z^{cx})$, 
where both the even components $Z^x$ and the odd ones $Z^{cx}$ are chiral multiplets. 
The even $N=1$ multiplets leading to light modes 
in the 4D effective theory consist then of a minimal universal set containing the gravitational 
multiplet $E$, the dilaton $S$ and the universal overall K\"ahler modulus $T^0$, plus a 
variable non-universal set containing the $h_{1,1}-1 \ge 0$ relative K\"ahler moduli 
$T^a$ and the $h_{2,1} \ge 0$ complex structure moduli $Z^x$.
Note that the first set of fields come from the ${\bf 1}$ of $SU(3)$, whereas those 
in the second set come from the projection of the ${\bf 8}$ and ${\bf 6}$ of $SU(3)$,
which depends on the orbifold action.

The quantum consistency of the original 11D theory compactified on $S^1/Z_2$ requires the further 
presence of two sets of 248 vector multiplets forming the adjoint of $E_8^{\rm v} \times E_8^{\rm h}$, 
distributed at the two 10D fixed hyperplanes of $S^1/Z_2$ and involving some gauge bosons 
$A^{\rm v}_K, A^{\rm h}_K$. When compactifying on $T^6/\Gamma$, these fields decompose as 
$A^{\rm v,h}_K \to A^{\rm v,h}_\mu, A^{\rm v,h}_m$. For the internal symmetry, one has as before
$SO(6) \to U(1) \times SU(3)$ and $SU(3) \to G$. From each $A^{\rm v,h}_K$ we thus get 1 vector 
from $A^{\rm v,h}_\mu$ in the ${\bf 1}$ and 6 scalars from 
$A^{\rm v,h}_m \Leftrightarrow A^{\rm v,h}_i,A^{\rm v,h}_{\bar \imath}$ in the ${\bf 6} \to {\bf 3} \oplus {\bf \bar 3}$. 
In terms of 4D $N=1$ multiplets, this corresponds to 1 vector multiplet $V^{\rm v,h}$ and 3 chiral multiplets 
$Q^{\rm v,h}_i$. For each group generator, either the former, or latter, or none of these components can be preserved. 
We therefore finally get some adjoint vector multiplets $V^{\rm v,h}$ plus some charged chiral multiplets $Q_i^{\rm v,h}$.
From the perspective of the 5D theory obtained by first compactifying on $T^6/\Gamma$, these can be though as 
living on the 4D fixed hyper-planes of $S^1/Z_2$. The precise representation content depends 
on how the orbifold action is embedded into the gauge group. But a completely generic fact is that one naturally 
gets two sets of at most $3$ generations of charged chiral multiplets $Q_i^{\rm v}$ and $Q_i^{\rm h}$ in the 
visible and hidden sectors. In the following, we shall then drop the detailed index structure concerning 
the gauge group, and treat this sector in a universal way. 
We shall also use the index ${\rm s} = {\rm v,h}$ to label the two sectors. It should finally be mentioned 
that in addition to the above matter chiral multiplets coming from the bulk of $T^6/\Gamma$, consistency 
also requires further matter chiral multiplets at the fixed points of $\Gamma$. We shall however ignore 
these fields and focus on the former, for simplicity.

In practice, there turn out to be a very limited set of qualitatively distinct models that can be achieved 
with Abelian orbifolds ($\Gamma = Z_N, Z_N \times Z_M$), at least if one focuses on the moduli structure
and not on the gauge structure. These correspond essentially to the three possible maximal-rank subgroups 
$G$ of $SU(3)$:
\bea
\a\a G = SU(3) \,,\hspace{42pt} h_{1,1} = 9 \,,\;\; h_{2,1} = 0 \nn \\
\a\a G = SU(2) \times U(1) \,,\;\; h_{1,1} = 5 \,,\;\; h_{2,1} = 0,1 \label{3cases} \\
\a\a G = U(1) \times U(1) \,,\hspace{13pt} h_{1,1} = 3 \,,\;\; h_{2,1} = 0,1,3 \nn
\eea
In the following, for simplicity we shall restrict to models with $h_{2,1} = 0$, where 
no complex structure moduli arise. We furthermore notice that the number $h_{1,1}-1$ 
of relative K\"ahler moduli $T^a$ coincides with the dimension of the adjoint 
representation of $G$. In fact, one can check that these indeed transform in the 
adjoint of $G$. On the other hand, $T^0$ corresponds to a singlet of $G$.
All together, the $h_{1,1}$ moduli fields transform in the adjoint of $G \times U(1)$, 
which is a subgroup of $U(3)$. They can then be naturally described also in terms 
of $h_{1,1}$ K\"ahler moduli denoted by $T_{ij}$ and suitably defined out of $T^0$ 
and the $T^a$'s. In addition, we always have the dilaton $S$ and the matter and 
gauge fields $Q_i^{\rm s}$ and $V^{\rm s}$ from the two branes.

Due to the special role played by the group $U(3) = U(1) \times SU(3)$, it is convenient 
to recall some properties of the defining representation of its algebra. In this representation,
the $9$ generators $\lambda^A$, $A=0,1,\cdots,8$ are given by 1 $U(1)$ generator 
$\lambda^0$ proportional to the identity matrix and 8 $SU(3)$ generators 
$\lambda^a$, $a=1,\cdots,8$, proportional to the Gell-Mann matrices:
\bea
\a\a \lambda^0 = \frac 1{\sqrt{3}}
\left(\begin{matrix}
1 & 0 & 0 \cr
0 & 1 & 0 \cr
0 & 0 & 1
\end{matrix}\right) \,,\;\;
\lambda^1 = \frac 1{\sqrt{2}}
\left(\begin{matrix}
0 & 1 & 0 \cr
1 & 0 & 0 \cr
0 & 0 & 0
\end{matrix}\right)
\,,\;\;
\lambda^2 = \frac 1{\sqrt{2}}
\left(\begin{matrix}
0 & i & 0 \cr
\!\mbox{-}i\! & 0 & 0 \cr
0 & 0 & 0
\end{matrix}\right) \,, \nn \\
\a\a \lambda^3 = \frac 1{\sqrt{2}}
\left(\begin{matrix}
1 & 0 & 0 \cr
0 & \!\mbox{-}1\!  & 0 \cr
0 & 0 & 0
\end{matrix}\right) \,,\;\;
\lambda^4 = \frac 1{\sqrt{2}}
\left(\begin{matrix}
0 & 0 & 1 \cr
0 & 0 & 0 \cr
1 & 0 & 0
\end{matrix}\right)
\,,\;\;
\lambda^5 = \frac 1{\sqrt{2}}
\left(\begin{matrix}
0 & 0 & i \cr
0 & 0 & 0 \cr
\!\mbox{-}i\! & 0 & 0
\end{matrix}\right) \,, \\
\a\a \lambda^6 = \frac 1{\sqrt{2}}
\left(\begin{matrix}
0 & 0 & 0 \cr
0 & 0 & 1 \cr
0 & 1 & 0
\end{matrix}\right) \,,\;\;
\lambda^7 = \frac 1{\sqrt{2}}
\left(\begin{matrix}
0 & 0 & 0 \cr
0 & 0 & i \cr
0 & \!\mbox{-}i\! & 0
\end{matrix}\right)
\,,\;\;
\lambda^8 = \frac 1{\sqrt{6}}
\left(\begin{matrix}
1 & 0 & 0 \cr
0 & 1 & 0 \cr
0 & 0 & \!\mbox{-}2\! 
\end{matrix}\right) \,. \nn
\eea 
With the chosen normalization, we have $\Tr \, [\lambda^A] = \sqrt{3} \, \delta^{A0}$ and
\be
{\rm Tr} \, [\lambda^A \lambda^B] = \delta^{AB} \,,\;\;
{\rm Tr} \, [\lambda^a \lambda^b] = \delta^{ab} \,.
\ee
Moreover, the following completeness relations hold true:
\be
\lambda^A_{ij} \lambda^A_{pq} = \delta_{iq} \delta_{jp} \,,\;\;
\lambda^a_{ij} \lambda^a_{pq} = \delta_{iq} \delta_{jp} - \frac 13 \delta_{ij} \delta_{pq} \,. \label{rel}
\ee
It is now clear that there are two particularly convenient linear bases of K\"ahler moduli, $T_{ij}$ and $T^A$,
which can be related by the matrices $\lambda^A_{ij}$ as follows:
\be
T_{ij} = \lambda^A_{ij} T^A \;\Leftrightarrow\; T^A = \lambda^A_{pq} T_{qp} \,.
\ee 
This linear transformation on the K\"ahler moduli is the one that allows to directly relate the 4D effective 
theory of heterotic M-theory orbifolds, to that of ordinary perturbative heterotic orbifolds \cite{HetOrb}. 
More precisely, the $T^A$'s are the natural parametrization emerging when one first reduces from 11D 
to 5D and then from 5D to 4D. On the contrary, the $T_{ij}$'s are the natural parametrization emerging 
when one first reduces from 11D to 10D and then from 10D to 4D. This is related to the well know fact 
that in the first case $S$ describes the overall size of $T^6/\Gamma$, and the K\"ahler moduli are then 
split into $T^0$ describing the overall size of $S^1/Z_2$ and the $T^a$'s describing the relative sizes 
of $T^6/\Gamma$, whereas in the second case $S$ describes the overall size of $S^1/Z_2$, and the 
K\"ahler moduli $T_{ij}$ describing the overall and relative sizes of $T^6/\Gamma$  are then naturally 
treated all on the same footing.
  
To figure out what is the low-energy effective theory, one has to reduce the original 11D action down 
to 5D and finally to 4D, by retaining only the light zero modes. Comparing then with the general form 
of the Lagrangian for a 4D supergravity theory, one can deduce the superpotential $W^{\rm eff}$ and 
the K\"ahler potential $K^{\rm eff}$ \cite{DG,LLN,NOY,LOW,L}. The results coincide with those obtained in
weekly-coupled heterotic orbifolds \cite{STU1,STU2,STU3}. For $W^{\rm eff}$, one finds a cubic term 
of the form $W^{\rm eff} = c^{\rm v} \epsilon_{ijk} Q_i^{\rm v} Q_j^{\rm v} Q_k^{\rm v} 
+ c^{\rm h} \epsilon_{ijk} Q_i^{\rm h} Q_j^{\rm h} Q_k^{\rm h}$,
which comes from the original Yang-Mills interactions. This preserves both the $G \subset SU(3)$ flavor 
symmetry and the $U(1)$ $R$-symmetry. One may however imagine that additional terms breaking the latter
could arise from other dynamical effects, and promote this to a more general function
$W^{\rm eff} = f^{\rm v} \big(\epsilon_{ijk} Q_i^{\rm v} Q_j^{\rm v} Q_k^{\rm v}\big) + 
f^{\rm h} \big(\epsilon_{ijk} Q_i^{\rm h} Q_j^{\rm h} Q_k^{\rm h} \big)$, which preserves only the 
$G \subset SU(3)$ flavor symmetry. For $K^{\rm eff}$, one finds:
\be
K^{\rm eff} = - \log (S + \bar S) - \log Y \,,
\ee
where the quantity $Y$ is a function of the K\"ahler moduli and matter fields, given by
\be
Y = \det \big(J_{ij}\big) = \det \big(\lambda_{ij}^A J^A \big)
\ee
in terms of the combinations
\bea
\a\a J_{ij} = T_{ij} + \bar T_{ij} - \lambda^A_{ij} \, \bar Q^{\rm s}_m \lambda^A_{mn} Q^{\rm s}_n \,, \label{Jij} \\
\a\a J^A = T^A + \bar T^A - \bar Q^{\rm s}_m \lambda^A_{mn} Q^{\rm s}_n \label{JA} \,.
\eea
The cubic polynomial $Y$ can be rewritten more explicitly as
\be
Y = \frac 16\, d_{ijpqrs} J_{ij} J_{pq} J_{rs} = \frac 16\, d^{ABC} J^A J^B J^C \,,
\label{Ydef}
\ee
where the numbers $d_{ijpqrs}$ and $d^{ABC}$ are related by
\be
d_{ijpqrs} = \lambda^A_{ji} \lambda^B_{qp} \lambda^C_{sr}\, d^{ABC} \,,
\ee
and given by the following expressions:
\bea
\a\a d_{ijpqrs} = \epsilon_{ipr} \epsilon_{jqs} \,, \\
\a\a d^{ABC} = \epsilon_{ipr} \epsilon_{jqs} \lambda^A_{ij} \lambda^B_{pq} \lambda^C_{rs} \nn \\
\a\a \hspace{29pt}
= 2 \, \Tr\, [\lambda^{(A} \lambda^B \lambda^{C)}] - 3\, \Tr\,[\lambda^{(A}] \, \Tr\,[\lambda^B \lambda^{C)}] 
+ \Tr\,[\lambda^{(A}] \, \Tr\,[\lambda^B] \, \Tr\,[\lambda^{C)}] \,.
\eea
The above formulae are valid in all the 3 cases listed in (\ref{3cases}), with the understanding that the number of 
K\"ahler moduli and the allowed values for $A$ and $i,j$ should be suitably restricted. In the case 
$G= SU(3)$ and $h_{1,1} = 9$, one has all the $9$ fields $T_{ij}$, $i,j=1,2,3$, corresponding to $T^A$ 
with $A=0,\cdots,8$. In the case $G= SU(2) \times U(1)$ and $h_{1,1} = 5$, one only has the $5$ fields 
$T_{11},T_{12},T_{21},T_{22},T_{33}$, corresponding to $T^A$ with $A=0,1,2,3,8$. 
Finally, in the case $G= U(1) \times U(1)$ and $h_{1,1} = 3$, one only has the $3$ fields $T_{11},T_{22},T_{33}$, 
corresponding to $T^A$ with $A=0,3,8$. In these three different distinct cases, the polynomial $Y$ takes 
the following more explicit form in terms of $M_{ij} = T_{ij} + \bar T_{ij} - Q_i^{\rm s} \bar Q_{\bar \jmath}^{\rm s}$:
\bea
\a\a Y_1 = \det_{i,j=1,2,3} \big(M_{ij} \big) \,,\hspace{15pt} 
d_1^{ABC} = d_{SU(3) \times U(1)}^{ABC} \,,\hspace{30pt}
d_{1\,ijpqrs} = \epsilon_{ipr} \epsilon_{jqs} \,, \label{Y1} \\[0.5mm]
\a\a Y_2 = \det_{\underline{i},\underline{j}=1,2} \big(M_{\underline{i}\underline{j}} \big) M_{33} \,,\;\; 
d_2^{ABC} = d_{SU(2) \times U(1) \times U(1)}^{ABC} \,,\;\;
d_{2\,\underline{i} \underline{j} \underline{p} \underline{q} 33} 
= \epsilon_{\underline{i} \underline{p} 3} \epsilon_{\underline{j} \underline{q} 3} \,, \label{Y2} \\
\a\a Y_3 = M_{11} M_{22} M_{33}  \,,\hspace{28pt}
d_3^{ABC} = d_{U(1) \times U(1) \times U(1)}^{ABC} \,,\hspace{10pt}
d_{3\,112233} = \epsilon_{123} \epsilon_{123} \,. \label{Y3}
\eea
In the following, it will be convenient to rewrite these results by distinguishing more explicitly the
$A=0$ and the $A=a$ generators, which correspond respectively to the $U(1)$ R-symmetry
and the $G \subset SU(3)$ flavor symmetry. Notice first that from the properties of $\lambda^0$ 
and $\lambda^a$ it follows that
\be
d^{000} = \frac 2{\sqrt{3}}\,,\;\;
d^{00a} = 0 \,,\;\;
d^{0ab} = - \frac 1{\sqrt{3}} \, \delta^{ab} \,,\;\;
d^{abc} = 2\, \Tr[\lambda^{(a} \lambda^b \lambda^{c)}] \,. \label{dabc}
\ee 
Since $\lambda^0_{ij} = \delta_{ij}/\sqrt{3}$, it is moreover convenient to define 
\be
T = \frac 1{\sqrt{3}} \, T^0 = \frac 13 \big(T_{11} + T_{22} + T_{33} \big) \,.
\ee
One then finds:
\be
Y = J^3 - \frac 12 \, J \, J^a J^a + \frac 16 \, d^{abc} J^a J^b J^c \,,\;\; 
d^{abc} \; \mbox{of $G \subset SU(3)$} \,,
\ee
where $J = J^0/\sqrt{3}$ and $J^a$ are given by 
\bea
\a\a J = T + \bar T - \frac 13 \bar Q_i^{\rm s} Q_i^{\rm s} \,, \\
\a\a J^a = T^a + \bar T^a - \bar Q_i^{\rm s} \lambda^a_{ij} Q_j^{\rm s} \,.
\label{Ja}
\eea

One of our aims is to compare this expression to that obtained by compactifying pure 5D supergravity 
on $S^1/Z_2$. In that case, there is no dilaton modulus neither complex structure moduli, and there 
is only one radion K\"ahler modulus $T$. In addition, one may consider some chiral and vector multiplets 
$Q_i^{\rm v,h}$  and $V^{\rm h,v}$ at the two 4D fixed planes. The superpotential $W^{\rm eff}$ can be an arbitrary 
separable function of the fields $Q_i^{\rm v}$ and $Q_i^{\rm h}$: $W^{\rm eff} = f^{\rm v}(Q_i^{\rm v}) + f^{\rm h}(Q_i^{\rm h})$.
The K\"ahler potential $K^{\rm eff}$ takes instead the simple form \cite{seq,LS}
\be
K^{\rm eff} = - \log Y \,,
\ee
where
\be
Y = J^3 \,,
\ee
in terms of the combination
\be
J =  T + \bar T - \frac 13 \bar Q_i^{\rm s} Q_i^{\rm s} \,.
\ee

The particular way in which we have rewritten the above 4D effective K\"ahler potentials, 
by splitting the moduli into a universal combination $T$ related to the $U(1)$ $R$-symmetry 
and $h_{1,1}-1$ non-universal combinations $T^a$ related to the $G \subset SU(3)$ flavor 
symmetry, will allow us to identify very explicitly the contact terms that arise in models with 
non-universal moduli compared to the toy model with no such extra moduli. More precisely, 
we will look for these contact terms in the effective K\"ahler function 
$\Omega^{\rm eff} = -3\,e^{-K^{\rm eff}/3}$, which is
the natural substitute of $K^{\rm eff}$ in supergravity, and interpret them 
as contact terms of the type (\ref{contact}). We will also be able to verify that these 
contact terms can be understood by integrating out in a manifestly supersymmetric way, 
at the level of superfields, the heavy KK modes of the odd vector multiplets $V^a$ coming 
along with such moduli $T^a$. 

\section{Effective contact terms}

To compare the effective theory arising for heterotic M-theory orbifolds to that arising for 
the simple sequestered toy model, it is instructive to study the special limit in which $T + \bar T \gg 
T^a + \bar T^a, \bar Q_i^{\rm s} Q_j^{\rm s}$. This corresponds to taking the volume of 
$S^1/Z_2$ to be large, and is thus indeed appropriate for comparing heterotic M-theory 
models, viewed as motivated effective 5D  theories, to the toy example of sequestered model, 
which is defined directly in 5D. In this limit, which implies $J \gg J^a$, one can compute $Y^{1/3}$ 
as a power expansion. 
One finds:
\bea
Y^{1/3} \b=\b \Big(J^3 - \frac 12 \, J \, J^a J^a + \frac 16 \, d^{abc} \, J^a J^b J^c \Big)^{1/3}
= J \Big(1 - \frac 12 \frac {J^a J^a}{J^2} + \frac 16 \, d^{abc} \, \frac {J^a J^b J^c}{J^3} \Big)^{1/3} \nn \\ 
\b=\b J - \frac 16 \,\frac {J^a J^a}{J} + \frac 1{18} \, d^{abc} \, \frac {J^a J^b J^c}{J^2} + \cdots \,.
\eea
After expanding also the terms involving the matter fields in $J$, this yields the following expression 
for the effective K\"ahler function:
\bea
\Omega^{\rm eff} \b=\b - 3\, (S + \bar S)^{1/3} (T + \bar T) 
+ (S + \bar S)^{1/3} \bar Q_i^{\rm s} Q_i^{\rm s} \nn \\
\b\;\b \,+ \frac 12 \frac {(S + \bar S)^{1/3}\!}{T + \bar T} 
\big(T^a \!+ \bar T^a \!- \bar Q_i^{\rm s} \lambda^a_{ij} Q_j^{\rm s} \big)
\big(T^a \!+ \bar T^a \!- \bar Q_p^{\rm s'} \! \lambda^a_{pq} Q_q^{\rm s'} \big) \nn \\
\b\;\b \,+ \frac 16 \frac {(S + \bar S)^{1/3}\!}{(T + \bar T)^2} \Big[
\big(T^a \!+ \bar T^a \!- \bar Q_i^{\rm s} \lambda^a_{ij} Q_j^{\rm s} \big)
\big(T^a \!+ \bar T^a \!- \bar Q_p^{\rm s'} \lambda^a_{pq} Q_q^{\rm s'} \big) \, \bar Q_r^{\rm s''} \! Q_r^{\rm s''} \nn \\
\b\;\b \hspace{20pt} -\, d^{abc} \big(T^a\! + \bar T^a\! - \bar Q_i^{\rm s} \lambda^a_{ij} Q_j^{\rm s} \big)
\big(T^b\! + \bar T^b\! - \bar Q_p^{\rm s'}\! \lambda^b_{pq} Q_q^{\rm s'} \big)
\big(T^c\! + \bar T^c\! - \bar Q_r^{\rm s''}\! \lambda^c_{rt} Q_t^{\rm s''} \big) \Big] \nn \\
\b\;\b \, + \cdots \,. \label{nonseq}
\eea

In the simple sequestered model, on the other hand, one has $Y^{1/3} = J$ and the effective 
K\"ahler function takes the following very simple form:
\bea
\Omega^{\rm eff} \b=\b - 3\, (T + \bar T) + \bar Q_i^{\rm s} Q_i^{\rm s} \,. \label{seq}
\eea
This has the particularity of being sequestered, meaning that there is no contact 
term mixing the visible and the hidden brane fields. Actually, we see that there is 
also no direct coupling between these localized sector fields and the radion field. 
This results in vanishing soft terms for the $Q_i^{\rm v}$, even when both the 
$Q_i^{\rm h}$ and $T$ get auxiliary field VEVs.    

Comparing eq.~(\ref{nonseq}) to eq.~(\ref{seq}), we see that the former contains several new effects
related to the additional chiral multiplets $S$ and $T^a$, which come from an additional hypermultiplet 
${\cal H}=(S;S^c)$ and additional vector multiplets ${\cal V}^a = (T^a;V^a)$ in the 5D mother theory. 
But what is more, even when discarding these new fields by freezing them to the values $S=1/2$ and 
$T^a = 0$, which as we shall argue better below is a truncation that correctly describes the situation 
where these fields are supersymmetrically stabilized, there remain additional contact interactions 
mixing chiral multiplets of the visible and hidden sectors:
\bea
\Omega^{\rm eff} \b=\b - 3 \,(T + \bar T) + \bar Q_i^{\rm s} Q_i^{\rm s}
+ \frac 12 \frac 1{T + \bar T} \big(\bar Q_i^{\rm s} \lambda^a_{ij} Q_j^{\rm s} \big) 
\big(\bar Q_p^{\rm s'} \! \lambda^a_{pq} Q_q^{\rm s'} \big) \nn \\
\b\;\b \,+ \frac 16 \frac 1{(T + \bar T)^2} \Big[
\big(\bar Q_i^{\rm s} \lambda^a_{ij} Q_j^{\rm s} \big)
\big(\bar Q_p^{\rm s'} \lambda^a_{pq} Q_q^{\rm s'} \big) \, \bar Q_r^{\rm s''} \! Q_r^{\rm s''} \nn \\
\b\;\b \hspace{70pt} +\, d^{abc} \big(\bar Q_i^{\rm s} \lambda^a_{ij} Q_j^{\rm s} \big)
\big(\bar Q_p^{\rm s'}\! \lambda^b_{pq} Q_q^{\rm s'} \big)
\big(\bar Q_r^{\rm s''}\! \lambda^c_{rt} Q_t^{\rm s''} \big) \Big] + \cdots \,. 
\label{justQT}
\eea
The leading additional terms are clearly of the current-current type (\ref{contact}), with 
the identification $J_a^{\rm s} = - \bar Q_i^{\rm s} \lambda^a_{ij} Q_j^{\rm s}$. This reflects 
the fact that they originate from integrating out in a supersymmetric way the massive 
modes of the odd $N=1$ vector multiplets $V^a$ coming with the even chiral multiplets 
$T^a$ from 5D vector multiplets. We see that there are also subleading effects involving 
three and more currents, as expected on general grounds from the discussion of section 2. 
As we shall now see in some detail, it turns out that in order to understand the leading terms 
with two and three currents, a simple rigid supersymmetry treatment is sufficient, whereas 
to recover all the subleading terms, one needs to switch to a local supersymmetry 
treatment and keep track of gravitational corrections.

Let us consider first the leading additional terms with two and three currents. Treating $T$ as the 
radion field, and restricting to rigid supersymmetry, the action for the heavy 5D vector multiplets 
${\cal V}^a$ can be written in terms of the $N=1$ superfields $V^a$ and $T^a$ along the lines 
of \cite{MSS,AGW,MP}. The general structure of this action is compatible with that of 4D 
$N=2$ theories \cite{DGG}, the 5D origin forcing the prepotential to be at most cubic \cite{S}. 
More precisely, the action reads
\bea
{\cal L}_{\rm 5D}^{\rm rigid} \b=\b \int \! d^2 \theta 
\bigg[\frac 14 \,T\, {\cal F}_{ab} \Big(\frac {T^c}T\Big) \, W^{a\alpha} W^b_\alpha 
- \frac 1{48} \, {\cal F}_{abc} \, \bar D^2 \big(V^a {\raisebox{11pt}{$$}}^{\leftrightarrow} \hspace{-10pt}  
D^\alpha \partial_y V^b\big) W^c_\alpha \bigg] + {\rm c.c.} \nn \\
\b\;\b \,+ \int \! d^4 \theta \, (T + \bar T) \, {\cal F}\Big(\frac {J_y^a}{T + \bar T} \Big) + \cdots \,, \label{L5D}
\eea
with a cubic prepotential of the general form
\be
{\cal F}(Z^A) = \frac 12 \, Z^a Z^a - \frac 16 \, d^{abc} \, Z^a Z^b Z^c \,.
\ee
In the above expression and in the following ones, the dots refer to terms involving additional 
powers of $V^a$ that are required by gauge invariance for non-Abelian groups. Their detailed 
form has been studied in \cite{HEB,HMZ}. We do not write them explicitly, because they will not 
be really relevant for us, for the same reasons as in \cite{HMZ}. More precisely, as far as the 
dependence on the chiral multiplets is concerned, the only relevant modification arising in the 
non Abelian case is that one has to take the trace over the Lie-algebra indices. Since each 
$T^a$ comes with a matrix $\lambda^a$, we then simply find that the quadratic part involves 
${\rm tr}[\lambda^a \lambda^b] = \delta^{ab}$ whereas the cubic part involves 
${\rm tr}[\lambda^a \lambda^b \lambda^c] = \frac 12 d^{abc}$, exactly as in the Abelian case. 
The quantity $W^{a \alpha}$ is the usual super field-strength associated to $V^a$, whereas 
$J_y^a$ is a current defined as
\be
J_y^a = - \partial_y V^a + T^a + \bar T^a
- \bar Q_i^{\rm s} \lambda^a_{ij} Q_j^{\rm s} \, \delta(y - y_0^{\rm s}) + \cdots \,.
\ee
The $T^a \!+ \bar T^a\!$ term in this current is standard and completely determined by the 5D gauge invariance. 
The additional term depending on the localized matter fields is instead peculiar of the situation at hand. As also
noticed in \cite{PS}, its presence is dictated by the fact that the 5D vector fields $A_M^a = C_{Mi\bar \jmath} \lambda_{ji}^a$ 
inherit a non-trivially modified Bianchi identity from the one of the 3-form $C_{ABC}$ in the original 11D theory,
which involves the localized matter scalar fields $q_i^{\rm s} = A_i^{\rm s}$. 
Indeed, with the above definition the $\theta \sigma^\mu \bar \theta$ component of $J_y^a$ correctly reproduces 
the modified version of the mixed components of the field strength, namely $F^a_{\mu y} = \partial_\mu A_y^a - \partial_y A_\mu^a  
- i q_i^{{\rm s}*} \lambda^a_{ij} \, {\raisebox{11pt}{$$}}^{\leftrightarrow} \hspace{-10pt} \partial_\mu \, q^{\rm s}_j 
\delta(y- y_0^{\rm s}) + \cdots $. This follows from the fact that ${\rm Im}\,T^a \propto C_{y i \bar \jmath} \lambda^a_{ji}$.
On the other hand, the lowest component of $J_y^a$ must not contain any contribution localized on the branes, and 
should simply give $g_{i \bar \jmath} \lambda^a_{ji}$. In order for this to happen, we must define
${\rm Re}\, T^a \propto g_{i \bar \jmath} \lambda^a_{ji} - \frac 12 q_i^{{\rm s}*} \lambda^a_{ij} q^{\rm s}_j \delta(y - y_0^{\rm s})$.
The need for this non-trivial definition of the chiral multiplets is dictated by supersymmetry. It is well known to emerge 
also in the derivation of the 4D effective theory based on the matching of kinetic terms, due to the fact that the modification 
of the Bianchi identity induces a non-trivial shift involving the matter fields only in the kinetic terms of the pseudoscalars 
arising from $C_{ABC}$ and not in those of the scalars arising from $g_{AB}$. Here we see that the same phenomenon 
also emerges very clearly at the level of superfields, in the intermediate theory where the odd vector multiplets have 
not yet been integrated out. As a last consistency check, notice that the localized shift in the definition of $T^a$ does 
not affect the first term in (\ref{L5D}), which controls the part of the kinetic terms and the Chern-Simons terms for the 
odd vector multiplets that involves $F^a_{\mu \nu}$, since this vanishes at the two branes.

Since the vector superfields $V^a$ contain only massive KK modes, they can be integrated out in a manifestly 
supersymmetric way to determine the 4D low-energy effective theory below the compactification scale. This is done 
by dropping the first line of the Lagrangian (\ref{L5D}), which contains 4D space-time derivatives that can be neglected, and 
then by varying with respect to $V^a$. The resulting equation of motion is solved by setting the 5D currents $J_y^a$
to their 4D zero modes $J^a$, given by (\ref{Ja}).\footnote{Our normalization is such that the integral 
of a 5D field yields  the 4D field describing its zero mode.} 
Plugging back into the action, we find then:
\be
{\cal L}_{\rm 4D}^{\rm rigid} = \int \! d^4 \theta \, (T + \bar T) \, {\cal F}\Big(\frac {J^a}{T + \bar T} \Big) 
=  \int \! d^4 \theta \, \bigg[\frac 12\, \frac {J^a J^a}{T + \bar T} - \frac 16 \, d^{abc} \, \frac {J^a J^b J^c}{(T + \bar T)^2} \bigg]\,,
\ee
where 
\be
J^a = T^a + \bar T^a - \bar Q_i^{\rm s} \lambda^a_{ij} Q_j^{\rm s} \,.
\ee
This clearly reproduces the leading additional terms with two and three currents arising in the 4D effective theory, 
and explains their origin from the 5D viewpoint, provided $d^{abc}$ is identified with the one of (\ref{dabc}).
   
The above comparison can be generalized by including also the subleading terms involving more 
inverse powers of $T + \bar T$. Such terms are however genuine supergravity effects, and to keep track of them,
one needs to use an off-shell description of the 5D supergravity theory, where half of the supersymmetry 
is manifest. The required formalism has been developed in \cite{KO,FKO} and further elaborated in 
\cite{PST,PS,AS}. One distinctive feature is that the graviphoton is described on the same footing as 
the other odd gauge fields, through vector multiplets $V^A$, where $A=0,a$. Correspondingly, the 
$R$-symmetry current to which the graviphoton couples is treated on the same footing as the other 
flavor currents, and all together they are denoted by $J^A$, where $A=0,a$.
The correct supergravity completion of (\ref{L5D}) turns out to be
\bea
{\cal L}_{\rm 5D}^{\rm local} \b=\b \int \! d^2 \theta \, 
\bigg[\!-\! \frac 14 \,{\cal N}_{AB} (T^A)\, W^{A\alpha} W^B_\alpha
+ \frac 1{48} \, {\cal N}_{ABC} \, \bar D^2 \big(V^A {\raisebox{11pt}{$$}}^{\leftrightarrow} \hspace{-10pt}  
D^\alpha \partial_y V^B\big) W^C_\alpha \bigg] + {\rm c.c.} \nn \\
\b\;\b \,+ \int \! d^4 \theta \, (- 3)\, {\cal N}^{1/3}(J_y^A) + \cdots \,. \label{L5Dsugra}
\eea
In this expression, the rigid prepotential ${\cal F}$ has been substituted by the norm function 
${\cal N}$, which is a also a cubic polynomial, but homogeneous and depending on one more 
variable, of the form:
\be
{\cal N}(Z^A) = \frac 16 \, d^{ABC} Z^A Z^B Z^C \,.
\ee
The currents $J_y^A$ are instead defined as:
\bea
\a\a J_y^A = - \partial_y V^A + T^A + \bar T^A - \bar Q^{\rm s}_i \lambda^A_{ij} Q_j^{\rm s} \, \delta (y - y_0^s) + \cdots \,.
\eea
As before, the dots in the above expressions denote additional terms needed in the non-Abelian case,
which are however not relevant for our discussion.

In this case, integrating out the heavy vector multiplets is slightly less straightforward. The main 
reason for this is that there are some constraints implementing the fact that one of the vector multiplets
is not completely physical and must contain only a vector field, the graviphoton, but no scalar. 
Clearly, such additional vector multiplet cannot be integrated out as straightforwardly as the other 
vector multiplets, and one has to properly take into account the constraints. One way to do this at 
the superfield level was described in \cite{AS}. We will not discuss the details here, but just quote 
that the final result is simply the one that one may have naively expected, obtained by replacing all 
the currents with their zero modes in the term of the action that does not involve the vector fields. More
precisely, one finds
\be
{\cal L}_{\rm 4D}^{\rm local} = \int \! d^4 \theta \, (- 3)\, {\cal N}^{1/3}(J,J^a) \,,
\label{4Dapp}
\ee
where the currents $J = J^0/\sqrt{3}$ and $J^a$ are defined as before, namely
\bea
\a\a J = T + \bar T - \frac 13 \bar Q^{\rm s}_i Q_i^{\rm s} \,,\\
\a\a J^a = T^a + \bar T^a - \bar Q^{\rm s}_i \lambda^a_{ij} Q_j^{\rm s} \,.
\eea
It is now clear that this result manifestly reproduces the full dependence on the K\"ahler moduli
and matter fields in the 4D effective theory, provided we identify the coefficients $d^{ABC}$
appearing in the norm function with those defined in section 3, so that the norm function takes 
the following form:
\be
{\cal N}(Z,Z^a) = Z^3 - \frac 12\, Z \, Z^a Z^a + \frac 16\, d^{abc} \, Z^a Z^b Z^c \,.
\ee
This is in turn related to the prepotential introduced in the rigid limit:
\be
{\cal N}(Z,Z^a) = Z^3 - Z^3 \,{\cal F} \Big(\frac {Z^a}{Z} \Big) \,.
\ee
To make full contact with the rigid limit, one may study the limit $T^a \ll T$ and $J^a \ll J$.
Evaluating (\ref{4Dapp}), one reproduces then indeed the leading corrections with two 
and three currents, and actually also the leading term involving just the $R$-symmetry 
current, which is purely due to gravity.

The above analysis shows that the contact terms arising in the K\"ahler function of 
heterotic M-theory orbifolds, compared to the simple sequestered toy model, do 
indeed take the general form expected for the interactions induced by heavy vector 
fields. Their detailed form shows that subleading effects involving additional powers
of the hidden sector fields do indeed appear, as generically expected. We are now
in position to study the structure of the soft terms induced by these contact terms, 
and check whether subleading effects suppressed by the parameters (\ref{egau})
and (\ref{egra}) do arise or not.

\section{Soft scalar masses}
\setcounter{equation}{0}

In the string models we have considered in the previous section, the visible sector is constituted 
by the visible-brane fields, which we relabel here $Q^\alpha$ with $\alpha=1,2,3$, whereas the hidden 
sector may contain both the hidden-sector fields, which we shall relabel $X^i$ with $i=1,2,3$ to 
distinguish them more efficiently, and the moduli sector, which contains the dilaton $S$ and the 
K\"ahler moduli $T^A$ with $A=1,\cdots,h_{1,1}$. 
When supersymmetry is spontaneously broken in the hidden sector, soft supersymmetry
breaking terms are induced in the visible sector. In particular, the soft masses $m^2_{\alpha \bar \beta}$ 
of the visible sector scalar fields $Q^\alpha$ receive in general contributions coming from the VEVs of all the 
hidden sector auxiliary fields $F^S$, $F^A$ and $F^i$, and the values of these masses also 
depend on the VEVs of the hidden sector scalar fields $S$, $T^A$ and $X^i$. The values taken by the 
scalar and auxiliary fields of the hidden sector is model dependent, and the only model-independent constraints 
that can be put on these comes from the requirement that the supersymmetry breaking vacuum should have 
vanishingly small cosmological constant and be at least  metastable. This respectively fixes the length and 
constrains the direction of the vector of auxiliary field VEVs defining the Goldstino, once a given K\"ahler potential 
has been specified  \cite{GRS}. On the other hand, the structure of the scalar masses as functions of these hidden 
sector fields only depends on the structure of the K\"ahler potential, and more precisely on the direct couplings 
between visible and hidden sector fields that arise in the effective K\"ahler function $\Omega^{\rm eff}$. What 
matters for these is the geometry of the scalar manifold. More precisely, denoting collectively with indices 
$\Sigma=S,A,i$ the fields of the hidden sector, one finds:
\be
m^{2}_{\alpha \bar \beta} = - \Big(R_{\alpha \bar \beta \Sigma \bar \Theta} 
- \frac 13\, g_{\alpha \bar \beta}\, g_{\Sigma \bar \Theta} \Big) F^\Sigma \bar F^{\bar \Theta} \,,
\ee
the vanishing of the cosmological constant implying 
\be
g_{\Sigma \bar \Theta}\, F^\Sigma \bar F^{\bar \Theta} = 3\, m_{3/2}^2 \,.
\ee

For the models under discussion, the scalar geometry is of a very particular type. Recall that the K\"ahler 
potential takes the separated form
\be
K = - \log (S + \bar S) - \log Y(T^A,Q^\alpha,X^i) \,,
\ee
where the quantity $Y$ is a homogeneous cubic polynomial of the currents $J_{ij}$ or $J^A$ defined
in eqs.~(\ref{Jij}) and (\ref{JA}), which is specified by the numbers $d_{ijklpq}$ or $d^{ABC}$ as in 
eq.~(\ref{Ydef}). As a consequence, the scalar manifold ${\cal M}$ factorizes into the product of a universal
piece spanned by $S$ and a model-dependent piece spanned by $T^A,Q^\alpha,X^i$:
\be
{\cal M} = {\cal M}_S \times {\cal M}_Y \,.
\label{scalM}
\ee
For the universal part describing the dilaton, one finds:
\be
{\cal M}_S = \frac {SU(1,1)}{U(1)} \,.
\label{scalS}
\ee
This is a maximally symmetric coset space, with a curvature scale such that the following fixed-scale 
property is satisfied:
\be
K^S K_S = 1 \,.
\ee
For the other part, it turns out that in the three distinct models corresponding to flavor groups $G$ equal 
to $SU(3)$, $SU(2) \times U(1)$ and $U(1) \times U(1)$, which are defined by eqs.~(\ref{Y1})-(\ref{Y3}), 
one finds the following manifolds, with $n = n_Q + n_X$ denoting the total number of charged matter 
fields appearing in each of the three generations:
\bea
{\cal M}_{Y_1} \b=\b \frac {SU(3,3 + 3\, n)}{U(1) \times SU(3) \times SU(3 + 3\,n)} \,, \label{scalY1}\\
{\cal M}_{Y_2} \b=\b \frac {SU(2,2 + 2\,n)}{U(1) \times SU(2) \!\times\! SU(2 + 2\, n)} 
\times \frac {SU(1,1 + n)}{U(1) \times SU(1 + n)}\,, \\
{\cal M}_{Y_3} \b=\b \frac {SU(1,1 + n)}{U(1) \times SU(1 + n)} \times
\frac {SU(1,1 + n)}{U(1) \times SU(1 + n)} \times 
\frac {SU(1,1 + n)}{U(1) \times SU(1 + n)} \,. \label{scalY3}
\eea
These manifolds are very particular. First, they are symmetric cosets, and their 
Riemann tensor is therefore covariantly constant. Secondly, they are of 
the so-called no-scale type, meaning that if $I=A,\alpha,i$ denotes an index 
running over all the K\"ahler moduli $T^A$ and the visible and hidden brane
matter fields $Q^\alpha$ and $X^i$, one has $K_I K^I = 3$. To prove this, 
notice that the fact that $e^{-K} = Y$ is homogeneous of degree $3$ in the 
$J^A$'s implies that $K_A J^A = - 3$. Differentiating with respect to $T^A$, 
one finds that $K_{\bar B} + K_{A \bar B} J^A = 0$, whereas differentiating 
with respect to $Q^\alpha$, $X^i$ and observing that 
$K_A \partial_\alpha J^A = K_{\alpha}$, $K_A \partial_i J^A = K_i$, 
one deduces that $K_{\bar \beta} + K_{A \bar \beta} J^A = 0$, 
$K_{\bar \jmath} + K_{A \bar \jmath} J^A = 0$. Put together, these relations
imply then that $K_{\bar J} + K_{A \bar J} J^A = 0$. Acting now with the full 
inverse metric $K^{I \bar J}$, one concludes that
$K^I = - \delta^I_A J^A$, meaning that $K^A = - J^A$, $K^\alpha = 0$ and $K^i = 0$.
Finally, this implies that $K_I K^I = - K_A J^A$, and since $K_A J^A = - 3$
one finally gets as stated that
\be
K_I K^I = 3 \,,\;\; I = A,\alpha,i \,.
\label{nscond}
\ee
This no-scale property holds at any point, and implies further restrictions 
on the structure of the Riemann tensor for these spaces.\footnote{It is also
well-known that in the particular case where $n=0$, {\it i.e.} in the absence 
of matter fields, these manifolds are actually special-K\"ahler manifolds, 
implying an even simpler structure of the Riemann tensor.} It is a straightforward 
exercise to work out the details and express the various relevant quantities 
in terms of $Y$ (see for instance \cite{CGGLPS}. Using this quantity, the no-scale 
property implies that $Y_I Y^{I \bar J} Y_{\bar J} = 3/2\,Y$, where $Y^{I \bar J }$ 
denotes the inverse of the matrix $Y_{I \bar J}$. One also finds that 
$Y^I = 2\, Y\, Y^{I \bar J} Y_{\bar J}$ and $Y_I Y^I = 3\, Y^2$.
The metric and its inverse can then be written as follows:
\be
g_{I \bar J} = - \frac {Y_{I \bar J}}{Y} + \frac {Y_I Y_{\bar J}}{Y^2} \,,\;\;
g^{I \bar J} = - Y \,Y^{I \bar J} + 2\, Y^{I \bar N} Y_{\bar N} \, Y^{\bar J M} Y_{M} \,.
\ee
The Riemann tensor, on the other hand, takes the following form:
\bea
R_{I \bar J P \bar Q} \b=\b g_{I \bar J} \,g_{P \bar Q} + g_{I \bar Q} \,g_{P \bar J} 
- 2\, \bigg(\frac {Y_{IP}}Y - \frac {Y_{I P \bar S} \,Y^{\bar S M} Y_M}{Y} \bigg) 
\bigg(\frac {Y_{\bar J \bar Q}}Y - \frac {Y_{\bar J \bar Q R} \,Y^{R \bar N} Y_{\bar N}}{Y} \bigg) \nn \\
\b\;\b +\, \frac {Y_{I P \bar S} \, Y^{\bar S R} \, Y_{\bar J \bar Q R}}{Y} 
- \frac {Y_{I \bar J P \bar Q}}{Y} \,.
\eea

For comparison, and for later use, it is perhaps useful to recall at this stage that
the toy sequestered model corresponds to a scalar manifold which is identified
with
\be
{\cal M} = \frac {SU(1,1 + 3\, n)}{U(1) \times SU(1 + 3\, n)} \,.
\ee
Moreover, the curvature is such that the no-scale condition (\ref{nscond}) is satisfied.
This is therefore of the same type as the cosets (\ref{scalY1})--(\ref{scalY3}), but 
with the additional distinguishing property of being maximally symmetric. As a 
consequence of these properties, the Riemann tensor takes then the simple form
\be
R_{I \bar J P \bar Q} = \frac 13 \big( g_{I \bar J} \,g_{P \bar Q} + g_{I \bar Q} \,g_{P \bar J} \big) \,.
\ee
This directly implies that the soft scalar masses vanish identically, independently 
of the values of the hidden-sector auxiliary fields.

Using the above results, we can now write down a more explicit formula for the 
soft scalar masses in the models under consideration. Let us recall first that the 
visible sector fields $Q^\alpha$ have vanishing VEVs both for their scalar and 
auxiliary components. Let us moreover split the hidden sector fields into the 
dilaton $S$, and the K\"ahler moduli $T^A$ plus hidden brane fields $X^i$,
which we shall commonly label with a new index $\square = A,i$. One deduces 
then that the metrics controlling the kinetic terms of the visible and hidden sector 
scalar fields are given by:
\be
g_{\alpha \bar \beta} = - \frac {Y_{\alpha \bar \beta}}Y \,,\;\; g_{S \bar S} = \frac 1{(S + \bar S)^2} \,,\;\;
g_{\square \bar \triangle} = - \frac {Y_{\square \bar \triangle}}Y + \frac {Y_{\square} Y_{\bar \triangle}}{Y^2} \,.
\label{metr}
\ee
The soft masses acquired by the $Q^\alpha$'s are instead given by the following expression:
\bea
m^{2}_{\alpha \bar \beta} \b=\b - \frac 13 \, \frac {Y_{\alpha \bar \beta}}{Y} \frac {F^S \bar F^{\bar S}}{(S + \bar S)^2} \nn \\
\b\;\b - \bigg(\frac 23\, \frac {Y_{\alpha \bar \beta}Y_{\square \bar \triangle}}{Y^2} 
- \frac 23\, \frac {Y_{\alpha \bar \beta}Y_{\square} Y_{\bar \triangle}}{Y^3} 
+ \frac {Y_{\alpha \square \bar \delta} Y^{\bar \delta \gamma} Y_{\bar \beta \bar \triangle \gamma}}Y
- \frac {Y_{\alpha \bar \beta \square \bar \triangle}}Y \bigg) F^\square \bar F^{\bar \triangle}  \,.
\label{scal}
\eea
Finally, the condition of vanishing cosmological constant now reads:
\be
g_{S \bar S}\, F^S \bar F^{\bar S} + g_{\square \bar \triangle} F^\square \bar F^{\bar \triangle}= 3\, m_{3/2}^2 \,.
\ee

We are now ready to compute the physical soft scalar masses, obtained after suitably rescaling the fields 
around the vacuum in such a way to canonically normalize their kinetic terms. To illustrate the mechanism 
of mild sequestering, let us for a moment freeze the dilaton in a supersymmetric way to the reference value 
$S=1/2$, to get rid of it; we shall come back to its effects later on. Let us furthermore assume that the 
K\"ahler moduli $T^A$ are all stabilized in a supersymmetric way, with $F^A = 0$, whereas the hidden-brane 
fields are stabilized in a supersymmetry breaking way, with $F^i \neq 0$. We have then to evaluate the 
second line of (\ref{scal}), by specializing $\square,\triangle \to i,j$. This is in general a complicated function
of the scalar VEVs of $T^A$ and $X^i$. However, so are also the metrics (\ref{metr}), and the true dependence
on these scalar VEVs in the physical masses has thus two sources: the one from the bare mass terms and the 
one from the kinetic wave function factor. In order to work out this dependence, which as explained in section 2 
is one of the issues that we want to investigate more explicitly, we can however use the fact that the scalar manifold 
${\cal M}_Y$ turns out to have covariantly constant curvature. Since the metric is also covariantly constant, this implies 
that also the scalar masses are covariantly constant over the scalar manifold. This in turn implies that after properly 
rescaling the fields to canonically normalize their kinetic terms, the physical soft masses will not depend on the 
VEVs of the fields $T^A$ and $X^i$. Indeed, the required local field redefinition simply amounts to switching to 
normal coordinates around the given point, and in these coordinates covariant constancy becomes true 
constancy. We can thus evaluate the physical scalar masses at any point we want, since they do not depend
on the point. The most convenient choice is the point defined by $T^A = \sqrt{3}/2\,\delta^{A0}$, $X^i = 0$. Since 
$F^A = 0$, the equation $T^A = \sqrt{3}/2\,\delta^{A0}$ can be implemented at the superfield level, meaning that 
these fields can effectively be integrated out in a trivial way. It is easy to verify that this is indeed the case, by 
noticing that when $F^A = 0$, the only term in (\ref{scal}) that may be sensitive to the presence of the fields 
$T^A$ is the third one in the second line. But one needs to have $\gamma = A$ and/or $\delta = B$, and then the contribution 
vanishes because $Y_{\alpha \square \bar A}$ and $Y_{\bar \beta \bar \triangle B}$ are odd functions of the 
visible fields and have thus vanishing VEVs. On the other hand, the equation $X^i=0$ cannot be implement 
at the level of superfields, since $F^i \neq 0$, but one can nevertheless expand the superfield expressions 
around that point and keep only up to two more powers of the superfields $X^i$, which can be converted 
to auxiliary fields $F^i$. In practice, this means that we can take 
\be
Y = J^3 - \frac 12 \, J \, J^a J^a + \frac 16 \, d^{abc} J^a J^b J^c \,,
\ee
where now
\bea
\a\a J = 1 - \frac 13 \bar Q_\alpha Q_\alpha - \frac 13 \bar X_i X_i \,, \\
\a\a J^a = -\, \bar Q_\alpha \lambda^a_{\alpha \beta} Q_\beta - \bar X_i \lambda^a_{ij} X_j\,.
\eea
We can furthermore expand this expression and retain only terms which are at most quadratic in
each type of fields. This gives:
\bea
Y \b=\b 1 - \delta_{ij} \bar X_i X_j - \delta_{\alpha \beta} \bar Q_\alpha Q_\beta 
- \Big(\lambda^a_{\alpha \beta} \lambda^a_{ij} - \frac 23 \delta_{\alpha \beta} \delta_{ij} \Big) 
\bar Q_\alpha Q_\beta \bar X_i X_j + \cdots \,.
\eea
At the point of vanishing $Q^\alpha$ and $X^i$, one then finds $Y=1$, $Y_i = 0$, $Y_{\alpha \bar \beta} = - \delta_{\alpha \beta}$, 
$Y_{i \bar \jmath} = - \delta_{ij}$, $Y_{\alpha p \bar \delta} = 0$ and 
$Y_{\alpha \bar \beta i \bar \jmath} = - \lambda^a_{\alpha \beta} \lambda^a_{ij} + \frac 23\, \delta_{\alpha \beta} \delta_{ij}$.
Applying (\ref{scal}), this finally yields the following result for the physical soft scalar masses of the canonically
normalized visible sector fields, expressed in terms of the auxiliary fields of the canonically normalized 
hidden sector fields:
\be
\hat m^2_{\alpha \bar \beta} = - \lambda^a_{\alpha \beta} \lambda^a_{i j} \, \hat F^i \hat {\bar F}^{\bar \jmath} \,. 
\label{mres}
\ee
The VEVs of the auxiliary fields are arbitrary at this stage, except for the constraint arising from 
the vanishing of the cosmological constant, which implies:
\be
\delta_{i j} \, \hat F^i \hat {\bar F}^{\bar \jmath} = 3\, m_{3/2}^2 \,.
\ee

The above simple expression is the exact general form of the physical soft scalar masses, under the 
assumption that the only source of supersymmetry breaking comes from the hidden-brane fields. 
We see that although $\Omega^{\rm eff} = - 3\,Y^{1/3}$ has an infinite series of terms involving an
increasing number of currents, the physical scalar masses are really sensitive only to the term with 
two currents, due to the particular property that the manifold has covariantly constant curvature.
More technically speaking, both the metric and the scalar masses depend on the scalar VEVs, 
but when one locally switches to normal coordinates, any dependence on the VEVs disappears, 
because in these coordinates covariantly constant quantities become really constant. In these 
models, there are then no subleading corrections involving the parameter (\ref{egau}). To get 
completely convinced of this simple result, one may also evaluate more brutally the soft masses 
and then appropriately rescale the fields to canonically normalize the kinetic terms and get the 
physical masses, without using any of the above short-cuts, {\it i.e.} without truncating the 
K\"ahler moduli and without going to a particular point. This is done in appendix A for the 
three distinct cases that can occur, where it is shown that one recovers the same result 
(\ref{mres}), with the sum over the index $a$ suitably restricted to the relevant values 
for each model.

We are now ready to examine how the mechanism of mild sequestering may be implemented 
in these models. According to the discussion of section 2, we assume for this that the flavor 
symmetry $G$ is a global symmetry of the effective theory. Recall now that the global symmetry 
$G \subset SU(3)$ is linearly realized and is naturally defined on all the fields. The 
visible and hidden-brane fields $Q^\alpha$ and $X^i$ transform in the fundamental 
representation descending from the ${\bf 3}$ of $SU(3)$, with $k_a^\alpha = i \lambda^a_{\alpha \beta} Q^\beta$ 
and $k_a^{i} = i \lambda^{a}_{ij} X^j$. The K\"ahler moduli $T^{ij}$ or $T^b$ transform instead 
in the adjoint representation descending from the ${\bf 8}$ of $SU(3)$, with 
$k_a^{ij} = i \lambda^a_{ik} T^{kj} - i T^{ik} \lambda^a_{kj}$ or $k_a^b = f_{ac}^{\;\;\;b} T^c$. 
The effective K\"ahler potential is strictly invariant under the above global symmetry. One may now wonder 
whether the current $J_a$ introduced in section 3 coincide with the conserved current
implied by the global symmetry, at least in the limit of small values for the charged fields. 
But unfortunately this is not easy to check, since as already said there is no obvious simple 
superfield expression for these currents in supergravity. On the other hand, it does not seem 
to make much sense to apply the rigid supersymmetry formula, even in some approximation, 
since the K\"ahler potential of the effective theory, and in particular its dependence on the K\"ahler
moduli, strongly depend on gravitational effects. 

Let us now assume that only the hidden-brane fields $X^i$ have non vanishing VEVs for their 
auxiliary fields $F^i$, whereas the K\"ahler moduli are stabilized in a supersymmetric way.  
Due to the global symmetry, we know from the component field analysis done in section 2 that 
the $F^i$ satisfy the Ward identity (\ref{wardsugr}). Note that even if the symmetry acts on all the 
fields, this relation only concerns the hidden sector fields with non-vanishing $F^i$, {\it i.e.} the 
hidden-brane fields. To work out its implications on the soft terms (\ref{mres}), we then switch
to normal coordinates, in order to obtain the Ward identity for canonically normalized fields. 
One then gets $\hat \nabla_i \hat k_{a \bar \jmath} = i \lambda^a_{ji}$, and the quadratic constraint 
among the auxiliary fields becomes then:
\be
\lambda^a_{ji} \hat F^i \hat {\bar F}^{\bar \jmath} = 2\, \epsilon_a^{\rm gra} \, m_{3/2}^2\,.
\label{wardlin}
\ee
In this expression, which is written in units where $M_{\rm Pl} = 1$, the quantity 
$\epsilon_a^{\rm gra} $ is given by $\epsilon_a^{\rm gra} = {\rm Im}(\hat k_a^{i} \hat F_i)/m_{3/2}$.
But since in our case $K$ and $W$ are separately invariant, this can also be rewritten 
as $\epsilon_a^{\rm gra} = - {\rm Im}(\hat k_a^{i} \hat K_i)$. This expression does explicitly depend 
on the vacuum point. Notice however that it is proportional to $\hat {\bar X}^i \lambda^a_{ij} \hat X^j$, 
which is the square of an energy scale $v^a$ related to the breaking of the gauge 
symmetry, and it may thus be reasonably assumed to be small: $\epsilon_a^{\rm gra} \simeq 0$. 
Finally, using the Ward identity (\ref{wardlin}) in the result (\ref{mres}), one finds that the 
physical scalar masses do not exactly vanish, but are suppressed by the parameter 
(\ref{egra}): 
\be
\hat m^2_{\alpha \bar \beta} = - 2\, \epsilon_a^{\rm gra} \lambda^a_{\beta \alpha} \,m_{3/2}^2 \simeq 0 \,.
\ee

Summarizing, we find that for the models under consideration the mechanism of mild sequestering 
works as expected only approximately. The subleading corrections involving $\epsilon_a^{\rm gau}$ 
are absent, due to the very particular structure of the models, whereas those involving $\epsilon_a^{\rm gra}$
are present.  But this is not dramatic since these subleading corrections are naturally small. 

It is straightforward to generalize the above analysis to the interesting case where the dilaton also participates 
to supersymmetry breaking, whereas the K\"ahler moduli are still stabilized in a supersymmetric way. In this 
case, the physical soft scalar masses are found to be given by
\be
\hat m^2_{\alpha \bar \beta} = \frac 13 \, \delta_{\alpha \bar \beta} \, \hat F^S \hat {\bar F}^{\bar S} 
- \lambda^a_{\alpha \beta} \lambda^a_{i j} \, \hat F^i \hat {\bar F}^{\bar \jmath} \,. 
\ee
The condition for vanishing cosmological constant implies on the other hand that
\be
\hat F^S \hat {\bar F}^{\bar S} + \delta_{i j} \, \hat F^j \hat {\bar F}^{\bar \jmath} = 3\, m_{3/2}^2 \,.
\ee
Assuming now that $G$ is a good flavor symmetry of the hidden sector, and noting that the dilaton 
is inert under this symmetry, one finds as before the Ward identity:
\be
\lambda^a_{ji} \hat F^i \hat {\bar F}^{\bar \jmath} = 2\, \epsilon_a^{\rm gra} m_{3/2}^2 \simeq 0 \,.
\ee
It follows then that the soft terms are approximately universal and given by:
\be
\hat m^2_{\alpha \bar \beta} = \frac 13 \, \delta_{\alpha \bar \beta} \, \hat F^S \hat {\bar F}^{\bar S} 
- 2\, \epsilon_a^{\rm gra} \lambda^a_{\beta \alpha} \, m_{3/2}^2 \simeq 
\frac 13 \, \delta_{\alpha \bar \beta} \, \hat F^S \hat {\bar F}^{\bar S} \,.
\label{scaldil}
\ee

\section{Applications}

The mechanism of mild sequestering has several interesting applications in string models.
The perhaps most relevant of all has already been pointed out in \cite{KMS}, and consists in 
the possibility of changing the cosmological constant without affecting the soft scalar masses.
In other words, one may construct models where the hidden sector enjoys a splitting into two
subsectors, the first contributing both to the cosmological constant and to the soft scalar masses, 
and the latter only to the cosmological constant, with no significant effect on soft scalar masses. 
This splitting is enforced by a global symmetry, and is thus natural and controlled. It represents
then a robust realization of the general idea of uplifting, first proposed in \cite{KKLT}, in the
sense that the screening between the uplifting sector and the visible sector is enforced in 
a very transparent way.
What we would like to emphasize in this section is that for the same reasons, the mechanism 
of mild sequestering also offers a simple solution to the metastability problem affecting the 
scenario of dilaton domination, which preserves its virtue of yielding approximately 
flavor-universal soft terms. 

Let us recall the basic results of \cite{GRS} concerning the possibility of finding a metastable 
vacuum with broken supersymmetry and vanishing cosmological constant, applied to the 
hidden sector of the heterotic M-theory models considered in this paper. The main point is 
that metastability of the vacuum constrains also the direction along which the vector of auxiliary 
field VEVs defining the Goldstino is allowed to point, for a given form of the K\"ahler potential.
For the specific case of a scalar manifold of the type (\ref{scalM}), with the dilaton part taking 
the form (\ref{scalS}) and the remaining part being one of the no-scale coset manifolds 
(\ref{scalY1})--(\ref{scalY3}), the constraints implied by the vanishing of the cosmological 
constant and by metastability take the following form, in terms of canonically normalized fields:
\bea
\a\a \hat F^S \hat {\bar F}^{\bar S} 
+ \delta_{\square \triangle} \, \hat F^\square \hat {\bar F}^{\bar \triangle} = 3\, m_{3/2}^2 \,, \\
\a\a \big(\hat F^S \hat {\bar F}^{\bar S}\big)^2 
+ \frac 13 \big(\delta_{\square \triangle} \, \hat F^\square \hat {\bar F}^{\bar \triangle}\big)^2 < 3\, m_{3/2}^4 \,. 
\eea
These two relations imply that the ratio between $\hat F^S \hat {\bar F}^{\bar S}$ and 
$\delta_{\square \triangle} \, \hat F^\square \hat {\bar F}^{\bar \triangle}$ cannot be too big. 
More precisely, one finds:
\be
\frac {\hat F^S \hat {\bar F}^{\bar S}}
{\delta_{\square \triangle} \, \hat F^\square \hat {\bar F}^{\bar \triangle}} < 1 \,.
\ee
This means the the Goldstino direction specified by the vector of auxiliary field VEVs 
must point more along ${\cal M}_Y$ than ${\cal M}_S$, so that supersymmetry breaking 
from the $Y$ sector is always stronger than supersymmetry breaking from the $S$ sector.

At this point, one may consider the situation in which the only fields breaking 
supersymmetry in the $Y$ part are the hidden-brane fields, and then invoke the mild 
sequestering following from assuming the existence of the global symmetry $G$ in 
that sector. In such a situation, metastability still implies that 
$\delta_{i j} \hat F^i \hat {\bar F}^{\bar \jmath}$ is larger than 
$\hat F^S \hat {\bar F}^{\bar S}$. However, the scalar masses (\ref{scaldil}) are 
insensitive to the $\hat F^i$'s and only feel $\hat F^S$. As a result, they are dominantly induced 
by the dilaton, although the dilaton cannot dominate the whole process of supersymmetry breaking.

\section{Conclusions}
\setcounter{equation}{0}

In this paper, we have studied in some more detail how the mechanism of mild sequestering proposed
in \cite{KMS,SS} may be technically realized in supergravity and string models. The basic idea is that 
whenever the only contact terms mixing visible and hidden sector fields in the effective K\"ahler function 
are induced by integrating out heavy vector fields, scalar soft masses can be suppressed by assuming 
the existence of global symmetries in the hidden sector. We have however argued that in general one is left with two 
types of subleading contributions to the scalar masses, which are suppressed by the ratios between 
the scale $v$ of spontaneous breaking of the global symmetry in the hidden sector and respectively the 
vector mass scale $M$ and the Planck scale $M_{\rm Pl}$. The first effect is due to non-linear corrections 
to the simple current-current structure that one obtains when integrating out heavy vector superfields, 
whereas the latter is due to gravitational corrections to the Ward identities of the global symmetries. 
This has then motivated us to study in full detail the structure of soft scalar masses in some prototypical 
class of string models defined from M-theory, and the circumstances under which mild sequestering may 
be realized in these models.

To be as concrete as possible, we have studied the simplest and most tractable case of heterotic M-theory
compactified on orbifolds, for which the 4D effective theory is fully known. We have started by identifying 
and characterizing more explicitly the part of the effective K\"ahler function that is due to integrating out 
heavy vector multiplets. We have then focused our attention on the effects mixing the hidden-brane fields 
to the visible-brane fields, to analyze the circumstances under which it may display mild sequestering. 
We have found that in these particular models, the subleading effects due to non-linearities drop out, 
as a consequence of the very peculiar property that the scalar manifold is a coset space with covariantly 
constant curvature, whereas those due to gravity are present but naturally small. In the more general 
case of heterotic M-theory compactified on smooth Calabi-Yau manifolds, the 4D effective theory is 
only partly known, and the same analysis cannot be done as explicitly. However, the scalar manifolds 
are certainly no longer coset spaces, and we therefore expect that in this case both 
types of subleading effect will be present. We also expect that a very similar situation occurs in other
types of string models, the general lesson being that mild sequestering generically works only up 
to subleading effects, which can however be made small without any severe obstruction.

One may wonder how the idea of mild sequestering fits into the more general panorama of situations 
in which the hidden sector, or a subsector of it, does not lead to any soft scalar masses at the classical
level. The possibility of achieving this situation has been studied for instance in \cite{LPS}, with the aim 
of finding generalizations of the minimal no-scale situation \cite{noscale}. It was found that for a large class 
of coset scalar manifolds, including those appearing in the effective theory of heterotic M-theory orbifolds, 
it is indeed possible to make the soft scalar masses vanish classically, by suitably choosing the 
orientation of the Goldstino direction defining spontaneous supersymmetry breaking. More precisely, 
this means that given the K\"ahler potential $K$ defining the geometry, one has to restrict the superpotential 
$W$ defining the Goldstino direction in a very particular way, and more specifically in a way that clearly 
has to do with the isometry structure of the coset scalar manifold. In this framework, mild sequestering 
can then be viewed as a natural and robust motivation of having such a specific alignment of the 
Goldstino direction relative to the isometries, which is enforced by a global symmetry. Moreover, 
the fact that such an alignment is not needed in the case of the toy sequestered models is understood
as coming from the fact that in that case the scalar manifold is maximally symmetric.

\section*{Acknowledgments}

We thank J.~Louis and R.~Rattazzi for useful discussions. This work was 
supported by the Swiss National Science Foundation.

\appendix 

\section{Functional dependence of the scalar soft masses}

In this appendix, we summarize the direct computation of the physical soft scalar masses 
for the three distinct cases of heterotic M-theory orbifolds, with flavor group $G$ given 
respectively by $SU(3)$, $SU(2) \times U(1)$ and $U(1)\times U(1)$. We suppose that 
the dilaton $S$ and the K\"ahler moduli $T^A$'s are stabilized in a supersymmetric way, 
with $\langle S \rangle = 1/2$ and $\langle T^A \rangle = T^A$, whereas 
the hidden-brane fields $X^i$'s spontaneously break supersymmetry, with 
$\langle X^i \rangle = X^i + \theta^2 F^i$. The visible-brane fields, on the other hand, 
do not get any expectation value: $\langle Q^\alpha \rangle = 0$. Our main concern 
here is to verify explicitly that the dependence on $X^i$ cancels out from the physical 
scalar soft masses. A similar check could be done for the dependence on $T^A$, but
for simplicity we will set $T^A = \sqrt{3}/2\, \delta^{A0}$.

\subsection{$G = SU(3)$}

In this case, the fields are $Q^\alpha$ and $X^i$, with $\alpha,i=1,2,3$. The metrics and the scalar 
masses are found to be given by
\bea
\a\a g_{\alpha \bar \beta} = \Pi_{\alpha \beta} \,, \\
\a\a g_{i \bar \jmath} = \frac 1{1 - |X|^2} \, \Pi_{ij} \,,
\eea
and
\bea
m^2_{\alpha \bar \beta} \b=\b - \frac 1{1 - |X|^2} \, \Big(\Pi_{\alpha j} \Pi_{i \beta} 
- \frac 13\, \Pi_{\alpha \beta} \, \Pi_{i j} \Big) \, F^i \bar F^{\bar \jmath} \,,
\eea
where $|X|^2 = \bar X^i X^i = |X^1|^2 + |X^2|^2 + |X^3|^2$ and ($x=\alpha,i$):
\be
\Pi_{x y} = \delta_{xy} + \frac {\bar X^x X^y}{1 - |X|^2} = 
\frac 1{1 - |X|^2}\, \Pi^\parallel_{xy} + \Pi^\perp_{xy} \,.
\ee
The $3 \times 3$ matrix $\Pi_{xy}$ has one eigenvalue $1/(1-|X|^2)$ in the direction parallel to $\bar X^x$, and 
two eigenvalues $1$ in the directions orthogonal to $\bar X^x$. To canonically normalize the kinetic 
term, one can then define the new rescaled fields in the following way:
\bea
\a\a \hat Q^\parallel = \frac {Q^\parallel }{\sqrt{1-|X|^2}}\,,\;\;
\hat Q^\perp_{1,2} = Q^\perp_{1,2} \,, \\
\a\a \hat X^\parallel = \frac {X^\parallel}{1-|X|^2} \,,\;\;
\hat X^\perp_{1,2} = \frac {X^\perp_{1,2} }{\sqrt{1-|X|^2}} \,.
\eea
One then finds that the physical masses read:
\bea
\hat m^2_{\alpha \bar \beta} \b=\b - \Big(\delta_{\alpha j} \delta_{\beta i} 
- \frac 13\, \delta_{\alpha \beta} \delta_{i j} \Big) \hat F^i \hat {\bar F}^{\bar \jmath} \,.
\eea
Finally, this can be rewritten as
\bea
\hat m^2_{\alpha \bar \beta} \b=\b - \sum_{a=1}^8 \lambda^a_{\alpha \beta} \lambda^a_{ij} \, \hat F^i \hat {\bar F}^{\bar \jmath} \,.
\eea

\subsection{$G=SU(2) \times U(1)$}

In this case, the fields are $Q^{\underline{\alpha}}, Q^3$ and $X^{\underline{i}},X^3$, with 
$\underline{\alpha},\underline{i}=1,2$. The metrics and the scalar masses are found to be given by
\bea
\a\a g_{\underline{\alpha} \underline{\bar \beta}} = \Pi_{\underline{\alpha} \underline{\beta}} \,,\;\;
g_{3 \bar 3} = \frac 1{1 - |X^3|^2} \,, \\
\a\a g_{\underline{i} \underline{\bar \jmath}} = \frac 1{1 - |X|^2} \, \Pi_{\underline{i} \underline{j}} \,,\;\;
g_{3 \bar 3} = \frac 1{(1 - |X^3|^2)^2} \,,
\eea
and
\bea
m^2_{\underline{\alpha} \underline{\bar \beta}} \b=\b 
- \frac 1{1 - |X|^2} \, \Big(\Pi_{\underline{\alpha} \underline{j}} \Pi_{\underline{i} \underline{\beta}} 
- \frac 13\, \Pi_{\underline{\alpha} \underline{\beta}} \, \Pi_{\underline{i} \underline{j}} \Big) \, 
F^{\underline{i}} \bar F^{\underline{\bar \jmath}} 
+ \frac 13\, \frac 1{(1 - |X^3|^2)^2} \, \Pi_{\underline{\alpha} \underline{\beta}} \, |F^3|^2 \,, \\
m^2_{3 \bar 3} \b=\b \frac 13\, \frac 1{1 - |X^3|^2}  \, \frac 1{1 - |X|^2} \, \Pi_{\underline{i} \underline{j}} \, 
F^{\underline{i}} \bar F^{\underline{\bar \jmath}}  - \frac 23 \, \frac 1{(1 - |X^3|^2)^3} \, |F^3|^2 \,,
\eea
where $|X|^2 = \bar X^{\underline{i}} X^{\underline{i}} = |X^1|^2 + |X^2|^2$ and ($\underline{x}=\underline{\alpha},\underline{i}$):
\be
\Pi_{\underline{x} \underline{y}} = \delta_{\underline{x} \underline{y}} + \frac {\bar X^{\underline{x}} X^{\underline{y}}}{1 - |X|^2} = 
\frac 1{1 - |X|^2}\, \Pi^\parallel_{\underline{x} \underline{y}} + \Pi^\perp_{\underline{x} \underline{y}} \,.
\ee
The $2 \times 2$ matrix $\Pi_{\underline{x} \underline{y}}$ has one eigenvalue $1/(1-|X|^2)$ in the direction parallel to
$\bar X^{\underline{x}}$, and one eigenvalues $1$ in the directions orthogonal to $\bar X^{\underline{x}}$. To canonically normalize 
the kinetic term, one can then define the new rescaled fields in the following way:
\bea
\a\a \hat Q^\parallel = \frac {Q^\parallel }{\sqrt{1-|X|^2}}\,,\;\;
\hat Q^\perp = Q^\perp \,,\;\; 
\hat Q^3 = \frac {Q^3}{\sqrt{1 - |X^3|^2}} \,, \\
\a\a \hat X^\parallel = \frac {X^\parallel}{1-|X|^2} \,,\;\;
\hat X^\perp = \frac {X^\perp}{\sqrt{1-|X|^2}} \,,\;\;
\hat X^3 = \frac {X^3}{1 - |X^3|^2} \,.
\eea
One then finds that the physical masses read 
\bea
\hat m^2_{\underline{\alpha} \underline{\bar \beta}} \b=\b 
- \Big(\delta_{\underline{\alpha} \underline{j}} \delta_{\underline{\beta} \underline{i}} 
- \frac 13\, \delta_{\underline{\alpha} \underline{\beta}} \delta_{\underline{i} \underline{j}} \Big) 
\hat F^{\underline{i}} \hat {\bar F}^{\underline{\bar \jmath}} 
+ \frac 13 \, \delta_{\underline{\alpha} \underline{\beta}} \, |\hat F^3|^2 \,, \\
\hat m^2_{3 \bar 3} \b=\b \frac 13\, \delta_{\underline{i} \underline{j}} \, 
\hat F^{\underline{i}} \hat {\bar F}^{\underline{\bar \jmath}} - \frac 23 \, |\hat F^3|^2 \,.
\eea
Finally, switching back to indices taking three values, $\alpha,i=1,2,3$, this 
can be rewritten more conveniently as:
\bea
\hat m^2_{\alpha \bar \beta} \b=\b 
- \!\! \sum_{a=1,2,3,8}\!\! \lambda^a_{\alpha \beta} \lambda^a_{ij} \, \hat F^i \hat {\bar F}^{\bar \jmath} \,.
\eea

\subsection{$G= U(1) \times U(1)$}

In this case, the fields are $Q^\alpha$ and $X^i$, with $\alpha,i=1,2,3$. The metrics 
and the scalar masses are found to be given by
\bea
\a\a g_{\alpha \bar \beta} = \frac 1{1 - |X^\alpha|^2} \,\delta_{\alpha \beta} \,, \\
\a\a g_{i \bar \jmath} = \frac 1{(1 - |X^i|^2)^2} \, \delta_{i j} \,,
\eea
and
\be
m^2_{\alpha \bar \beta} = - \frac {\delta_{\alpha \beta}}{1 - |X^\alpha|^2} 
\bigg(\frac {|F^\alpha|^2}{(1 - |X^\alpha|^2)^2} 
- \frac 13\, {\sum}_{\gamma} \, \frac {|F^\gamma|^2}{(1 - |X^\gamma|^2)^2} \bigg) \,.
\ee
One can now rescale the fields in the following way to canonically normalize their kinetic terms:
\bea
\a\a \hat Q^\alpha = \frac {Q^\alpha}{\sqrt{1 - |X^\alpha|^2}} \,, \\
\a\a \hat X^i = \frac {X^i}{1 - |X^i|^2} \,.
\eea
The physical soft masses are then found to be:
\bea
\hat m^2_{\alpha \bar \beta} \b=\b - \delta_{\alpha \beta} \, \Big(|\hat F^\alpha|^2
- \frac 13\, {\sum}_{\gamma} \, |\hat F^\gamma|^2 \Big) \,.
\eea
Finally, this can be rewritten as follows:
\bea
\hat m^2_{\alpha \bar \beta} \b=\b 
- \sum_{a=3,8} \lambda^a_{\alpha \beta} \lambda^a_{ij} \, \hat F^i \hat {\bar F}^{\bar \jmath} \,.
\eea

\end{document}